\documentclass[10pt]{article}

\usepackage[
    letterpaper,
    margin=1in
]{geometry}

\usepackage[T1]{fontenc}
\usepackage{newtxtext}
\usepackage[expansion=false]{microtype}

\usepackage{amsmath,amssymb}
\usepackage{bm}

\usepackage{booktabs}
\usepackage{multirow}
\usepackage{tabularx}
\usepackage{longtable}
\usepackage{threeparttable}
\usepackage{siunitx}
\usepackage{xcolor}

\usepackage{graphicx}
\usepackage{float}
\usepackage{placeins}
\usepackage{subcaption}
\usepackage{xcolor}

\definecolor{HumeInk}{HTML}{353535}
\definecolor{HumeBlue}{HTML}{6E9EE8}
\definecolor{HumeLightBlue}{HTML}{EEF3FB}
\definecolor{HumeRule}{HTML}{E6E6E1}
\definecolor{HumeGray}{HTML}{6B6B6B}

\usepackage{enumitem}

\usepackage[numbers,sort&compress]{natbib}
\usepackage[
    colorlinks=true,
    linkcolor=HumeBlue,
    citecolor=HumeBlue,
    urlcolor=HumeBlue,
    hyperfootnotes=false
]{hyperref}
\usepackage[nameinlink,capitalise,noabbrev]{cleveref}

\usepackage{titlesec}
\usepackage{titling}
\usepackage[font=small,labelfont=bf,labelsep=period]{caption}
\usepackage[most]{tcolorbox}
\usepackage{comment}

\graphicspath{{figures/}}

\titleformat{\section}{\Large\bfseries\color{HumeInk}}{\thesection}{0.65em}{}
\titleformat{\subsection}{\large\bfseries\color{HumeInk}}{\thesubsection}{0.65em}{}
\titleformat{\subsubsection}{\normalsize\bfseries\color{HumeInk}}{\thesubsubsection}{0.65em}{}
\titlespacing*{\section}{0pt}{2.0ex plus 0.6ex minus 0.2ex}{0.8ex}
\titlespacing*{\subsection}{0pt}{1.5ex plus 0.4ex minus 0.2ex}{0.5ex}
\titlespacing*{\subsubsection}{0pt}{1.0ex plus 0.3ex minus 0.2ex}{0.35ex}

\newcommand{\humelogo}{%
  {%
    \includegraphics[width=1.5in]{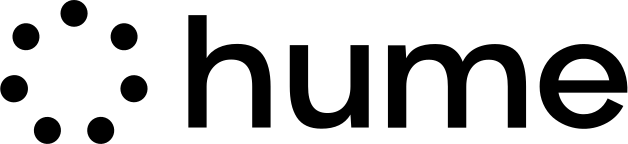}
  }
}
\pretitle{\noindent\humelogo\par\vspace{1.8em}\begin{center}\LARGE\bfseries\color{HumeInk}}
\posttitle{\par\vspace{0.55em}{\color{HumeRule}\rule{0.68\linewidth}{0.6pt}}\end{center}}
\preauthor{\begin{center}\normalsize\color{HumeGray}\begin{minipage}{0.92\linewidth}\centering}
\postauthor{\end{minipage}\end{center}\vspace{-0.6em}}
\predate{}
\postdate{}
\renewenvironment{abstract}{%
  \begin{center}
  \begin{tcolorbox}[enhanced,width=0.94\linewidth,colback=HumeLightBlue,colframe=HumeRule,boxrule=0.5pt,arc=1.5pt,left=9pt,right=9pt,top=7pt,bottom=7pt]
  \small\noindent\textbf{Abstract.}\space
}{%
  \end{tcolorbox}
  \end{center}
}

\AtBeginDocument{\color{HumeInk}}
\renewcommand{\arraystretch}{1.12}
\setlist[itemize]{leftmargin=2.5em, itemsep=0.25em, topsep=0.25em}

\title{Towards Quantifying Benchmark Optimization in ASR Models}
\author{
Theo Lebryk\thanks{Corresponding author: \texttt{theo@hume.ai}}, 
David Ayllon,
Alice Baird,
Jakub Piotr C\l{}apa,
Jens Madsen,
Panagiotis Tzirakis
\\[0.45em]Hume AI Research
}
\date{}

\begin{document}
\maketitle

\begin{abstract}
Public benchmarks are important measures of Automatic Speech Recognition (ASR) model capabilities.
However, by nature of being public, there is risk of models being optimized for these benchmarks in ways that do not generalize well to real-world data.
We present a methodology for quantifying benchmark optimization, focusing on cases where the audio underdetermines the reference transcript~\footnote{\href{https://github.com/HumeAI/asr-benchmark-optimization}
{\texttt{github.com/HumeAI/asr-benchmark-optimization}}.}.
We identify three families of behavioral probes that reveal models' capabilities of reproducing benchmark reference spans despite underdetermined audio: reference disagreement, masked-number recovery, and orthographic switching.
We find that the highest-scoring open source models output verbatim reference transcript spans even when the relevant audio is contradictory, masked, or ambiguous.
Using a variety of mechanistic probes, 
we show that models respond to narrow acoustic cues 
to override the faithful representation of the audio in favor of a benchmark-optimized policy. 
We show the benchmark-optimized behavior can be causally manipulated via low-rank linear steering or simply appending audio to the end of a segment in some cases.
Overall, our results indicate that high-performing models exhibit benchmark-conditioned behaviors that can inflate benchmark performance without reflecting improved general-purpose transcription ability.

\end{abstract}

\section{Introduction}
\label{sec:intro}

Researchers in Automatic Speech Recognition (ASR) models have claimed models have achieved human-level performance on public benchmarks for nearly a decade \citep{amodei2016deepspeech2,xiong2016human}, yet a
persistent gap separates benchmark performance from real-world utility \citep{radford2023whisper, xu2026megaasr}.
Ideally, a low word error rate (WER) on a benchmark reflects a general ability to transcribe speech that extends to unseen, real-world audio. Because benchmarks are public, however, models can be optimized to
drive their reported WER down in ways that are orthogonal to---or even actively harmful
to---real-world transcription abilities, causing benchmark scores to overstate a model's general-purpose performance.

We define benchmark optimization (colloquially known as benchmaxxing) as gains in reported performance that arise from reliance on benchmark-specific artifacts rather than from a generalizable improvement in transcription ability.

To measure this phenomenon, we construct cases in which the audio does not uniquely support the exact reference transcript. For instance, when a benchmark entry contains a transcription
error, an audio-faithful transcriber should place low probability on the (erroneous) reference transcript.
Similarly, if the audio for a word is masked, the model should also place limited probability on that word.
Finally, when a word admits two phonetically and semantically equivalent renderings, the model should not switch between them to match the benchmark's local reference convention.
In all these ``benchmark optimization probes,'' models systematically matching the reference transcript at rates far above chance suggests that the model is using the benchmark's acoustic cue as a shortcut
rather than faithfully listening to the audio. We find that these behaviors persist in synthetic speech using clones of the voices of speakers from the benchmark's evaluation set, but weaken for clones of generic speakers or previously unseen speakers from independently collected data in the same source domain.
Beyond establishing that benchmark optimization happens, we ask when and how this behavior is triggered. 

We find that models override audio-supported content at different depths, triggered by a fairly narrow range of inputs, and largely operate faithfully off the benchmark's distribution. However, steering a model's internal activations along a low-dimensional linear direction or appending benchmark-like or generic audio
can bi-directionally flip the benchmark-optimized behavior.
As such, models are expressive enough to localize the benchmark-conditioned representations to a narrow set of acoustic cues, limiting its effect off-distribution while inflating measured performance on the benchmark.
We use “narrow” to mean that the behavior is triggered by a restricted range of acoustic contexts: 
the benchmark-optimized policy appears on benchmark recordings and test-speaker clones, but often weakens on generic voices and newly collected speakers from the same domain.

We make the following contributions:
\begin{itemize}
  \item We present a reusable methodology for measuring ASR benchmark optimization
        from model behavior: reference disagreement, masked-number recovery, and orthographic switching.
  \item We show that benchmark-optimized behavior is common among the
   highest-scoring models on two public benchmarks, with models prone 
    to reproducing benchmark-specific transcripts despite audio evidence to the contrary.
  \item We show how benchmark-specific acoustic context changes model behavior at inference time.
  The model can faithfully transcribe the target speech when its context is restricted, 
  but activates a benchmark-optimized policy when presented with sufficient benchmark-specific cues. We further demonstrate two bidirectional interventions that can activate or suppress this policy.
\end{itemize}

\section{Related Work}
\label{sec:related}

It is well established that state-of-the-art ASR model WER on standard benchmarks often overstates real-world ability \citep{szymanski2020wer, likhomanenko2021rethinking, radford2023whisper}.
The distributional limits of industry-standard datasets have been studied and numerous recent works have proposed new robustness-oriented ASR evaluations targeting acoustic degradation \citep{shah2025srb,xu2026megaasr}, far-field \citep{dai2025aishell5}, and other real-world conditions \citep{ayllon2026rw, bezzam2026benchmaxxer, tay2026wildasr}.
However, this line of work primarily treats the benchmark--reality gap as a coverage problem: if existing benchmarks miss important conditions, add benchmarks that include them. Our results show that there is also a measurement problem.
Public benchmarks can reward benchmark-specific behavior even when the model is not faithfully transcribing the audio.
This issue is especially relevant for new benchmarks derived from synthetic augmentations of existing datasets \citep{goswami2026whisperrirmega,shah2025srb,xu2026megaasr}\, as
we find that several models' benchmark-specific behavior can persist under exactly these
perturbations (App.~Figure~\ref{fig:trig-robustness}).
New datasets are necessary but insufficient: without a framework to detect and measure benchmark optimization, new benchmarks may be subject to the same measurement distortions as existing evaluation sets.

Earlier ASR systems added bespoke text post-processing steps per benchmark to optimize performance on different
orthographic conventions, which can account for a significant portion of cross-corpus error \citep{likhomanenko2021rethinking};
reporting model performance across an array of benchmarks using a single text post-processing procedure has alleviated the use
of bespoke processing steps to benchmark-optimize \citep{gandhi2022esb, srivastav2025openasrleaderboardreproducible}.
However, this shift merely pushes the issue of benchmark optimizing using bespoke convention matching \citep{akeret2026swiss} to the model level.
A model which is expressive enough to learn the conventions of multiple benchmarks and alternate between conventions based on arbitrary acoustic cues
can still hill-climb on benchmarks independent of its general-purpose competency.

Deep learning models have traditionally been known to exploit statistical regularities in the training data as shortcuts to
completing a task in a variety of domains
\citep{jo2017measuring, poliak2018hypothesis,gururangan2018annotation,mccoy2019right, liu2024clever, geirhos2020shortcut}.
With the emergence of speech-large language models (LLMs) for ASR, there is evidence of evaluation set transcripts leaking into speech-LLM's decoder backbone \citep{tseng2025evaluation}.
However, this form of contamination is localized to a specific training paradigm, while our results show evidence of benchmark optimization across multiple architectures.
Moreover, authors have found the impact of decoder data contamination
on WER to be relatively small, reinforcing recent NLP research which suggests that ``training on the test task''
is a more prominent form of benchmark optimization than data contamination \citep{dominguezolmedo2024training}.
Thus, there is a need for an updated understanding and measurement framework for how models optimize performance on ASR benchmarks.

Researchers have begun to apply mechanistic interpretability techniques to
automatic speech recognition (ASR) \citep{pluth2026mechanistic,glazer2026beyond}.
We apply related interventions
to benchmark optimization, using context manipulation, activation patching \citep{zhang2024towards, vig2020investigating},
and activation steering \citep{turner2023steering, rimsky2024steering} to localize and causally manipulate benchmark-specific transcription behavior.

\section{Method}
\label{sec:method}

\subsection{Experimental setup}
\label{sec:setup}

\paragraph{Models.}
We evaluate 11 widely used open source ASR models spanning the two dominant architectures:
\emph{encoder--decoder} attention/transducer models (Whisper-Large-v3~\citep{radford2023whisper},
Cohere-Transcribe~\citep{cohere2026transcribe},
Parakeet-TDT-0.6B-v2~\citep{nvidia2025parakeet,xu2023tdt},
Moonshine-Streaming~\citep{jeffries2024moonshine}) and \emph{speech--LLM} models (Canary-Qwen-2.5B~\citep{nvidia2025canaryqwen,chen2024salm}, 
Granite-Speech-4.1-2B~\citep{saon2025granitespeech}, 
Higgs-Audio-v3-8B~\citep{bosonai2026higgsaudio}, 
Kimi-Audio-7B~\citep{kimiteam2025kimiaudio}, 
Phi-4-Multimodal~\citep{microsoft2025phi4},
Qwen3-ASR-0.6B~\citep{shi2026qwen3asr},
Voxtral-Mini-3B~\citep{mistral2025voxtral}).
We report teacher-forced likelihood metrics for all models but Parakeet-TDT, whose token-and-duration transducer architecture makes reading logits at arbitrary positions difficult.

\paragraph{Datasets.}
We focus our analysis on \textbf{VoxPopuli}~\citep{wang2021voxpopuli} (English), which
consists of European Parliamentary recordings and is among the most widely used ASR benchmarks.
The dataset has a train, validation, and test split but $40\%$ of speakers in the test split are leaked into the training split in the Hugging Face version of the dataset.
Where applicable, we extend our findings to \textbf{LibriSpeech}~\citep{panayotov2015librispeech} (\textsc{clean} and \textsc{other}).
As a held-out control, we use \textbf{\textsc{DaiKon}}, a private
set of $450$ conversational clips drawn from the naturalistic dyadic-conversation
collection of \citet{tzirakis2026daikon}; these clips postdate and lie outside every
model's training data and also point to how the capabilities measured in the public benchmarks
translate to real-world, conversational audio.
To probe the limits of when benchmark optimized behavior occurs, we add two further sources. We use
Qwen3-TTS~\citep{hu2026qwen3} to generate synthetic samples using a variety of reference speakers.
We also collect fresh data corresponding to a new held-out dataset from the same domain as
both benchmarks. 
We scrape European Parliament recordings from June
2026 (\textsc{ep-fresh})---after every model's training cutoff---following the original VoxPopuli
collection procedure. Manual inspection confirmed these transcript/clip pairs are in-distribution, down to characteristic
VoxPopuli reference errors such as courtesy-expression omission. 

For LibriSpeech, we analogously collect 2026 LibriVox recordings from $14$ newly active
readers whose catalog histories begin after every model's training cutoff
(\textsc{libri-fresh}).

\subsection{Probes and readouts}
\label{sec:method-framework}

Every behavioral probe marks a set of positions where the audio $x$ underdetermines the reference and
contrasts the \emph{reference rendering} $r$ with a competing \emph{audio-true} (or acoustically
equivalent) rendering $a$ (Figure~\ref{fig:probe-examples}). 
We start our analysis by looking only at the surface transcript produced by standard greedy decoding;
\textsc{accept-ref} is the fraction of these positions at which the model emits $r$ rather than
$a$. A transcriber faithful to the sound should follow $a$ at high rates, so a high \textsc{accept-ref} indicates the
model $M$ is reproducing the benchmark's reference beyond its acoustic content. Each probe in
Sections~\ref{sec:method-consensus}--\ref{sec:method-ortho} supplies its own $(r,a)$; we name the
per-probe instances \emph{reference-disagreement}, \emph{masked}, and \emph{orthographic}
\textsc{accept-ref}.

To isolate the effect of the raw language model prior compared to the end-to-end model,
we subtract the silenced-audio prior $x_\emptyset$ (the clip with its waveform
zeroed) from the teacher-forced log-likelihood of the reference span given the audio and the flanking transcript context. To account for different span lengths in a way that generalizes across tokenizers, we normalize by the character count of the span, $|r|_{char}$.
We define the \emph{audio lift} as:
\begin{equation}
  \lambda(r) \;=\; \frac{\log p_M\!\bigl(r\mid x\bigr) - \log p_M\!\bigl(r\mid x_\emptyset\bigr)}{|r|_{char}}.
  \label{eq:lift}
\end{equation}
$\lambda(r)>0$ means the audio, not the prior, raised the model's likelihood for the reference transcript.
A high audio lift suggests the model is relying on benchmark specific artifacts: We've curated cases explicitly where the surrounding audio
should underdetermine the reference transcript.

\begin{figure}[tbp]
\centering
\small
\setlength{\tabcolsep}{5pt}
\renewcommand{\arraystretch}{1.25}
\begin{tabular}{@{}p{0.22\linewidth}p{0.34\linewidth}p{0.34\linewidth}@{}}
\toprule
Probe case & Reference rendering $r$ & Comparison rendering $a$ \\
\midrule
Reference insertion & we should return to \colorbox{red!10}{the} Geneva format & we should return to Geneva format \\
Reference deletion & Mr. President & \colorbox{green!10}{Thank you} Mr. President \\
Reference substitution & in a \colorbox{red!10}{position}  & in a \colorbox{green!10}{role} \\
\addlinespace
Masked word & we also had \colorbox{red!10}{50} international US experts & we also had \colorbox{green!10}{\texttt{<masked>}} international US experts \\
Orthographic variant & \colorbox{red!10}{Mr} President & \colorbox{green!10}{Mister} President \\
\bottomrule
\end{tabular}
\caption{Example renderings used by the behavioral probes. Red marks the reference span being tested; green marks
the competing audio-true or acoustically equivalent rendering. Reference insertion, deletion, and
substitution are the three edit types used by the reference-disagreement probe. Masking induces a
deletion-style comparison by silencing a word (where \texttt{<masked>} indicates a silent audio interval, not a literal model output), while orthographic switching compares two acoustically
identical spellings.}
\label{fig:probe-examples}
\end{figure}

\subsection{Reference disagreement}
\label{sec:method-consensus}

A \emph{reference disagreement} is a span where the reference transcript contains an error,
meaning the audio contradicts the reference transcript.
These errors can include insertions (reference transcript added a word), omissions (reference transcript missed a word),
and substitutions (reference transcript transcribed a word as a different word).
Here $r$ is the erroneous reference span and $a$ is its audio-supported correction, so
\emph{reference-disagreement} \textsc{accept-ref} is the fraction of
reference errors for which the model reproduces the erroneous reference output rather than the
correction.

A model that has overfit to a benchmark will reproduce the erroneous $r$, lowering its WER against the flawed reference while
departing from what was spoken.
High reference-disagreement accept-ref therefore indicates that the model is using arbitrary cues from the benchmark to prefer the incorrect benchmark-optimal transcript over the audio-true one.

Reference disagreements can be mined from any source, ideally from multiple human labellers.
However, scaling human labellers can be costly and time-consuming.
To detect this at scale without human annotation of every clip, we use a
consensus panel of independent models to flag reference errors.
We select the panel using phoneme error rate (PER) so that its members are accurate transcribers.
We use Kimi-Audio, Qwen3-ASR-0.6B, Voxtral-Mini-3B, and Moonshine-Streaming for VoxPopuli.
For each clip we align every panel hypothesis to the reference and record per-word edits
that the audio supports against the reference. We treat edits the panel flags unanimously
as consensus-flagged reference errors ($1{,}113$ edits on $745$ VoxPopuli test clips:
$586$ substitutions, $441$ deletions, $86$ insertions). Panel members themselves are scored against edits flagged unanimously by the other
three (leave-one-out); otherwise, their \textsc{accept-ref} would be zero by construction.
Validated against a human-annotated subset \citep{artificialanalysis2026voxpopulicleaned}, $93\%$ of consensus-flagged edits also appear
in the human annotations (the human edits the consensus panel misses are predominantly formatting changes);
$40\%$ of all clips, and roughly $3\%$ of all reference words, carry a
flagged edit.

We focus on the VoxPopuli dataset for the reference disagreement probe as it is known to be rife with reference errors
and can be validated against human annotations,
but this consensus agreement approach can be extended
to datasets without existing human annotations
more generally.

\subsection{Masked-entity recovery}
\label{sec:method-masking}

The reference disagreement probe relies on naturally occurring reference errors. We can also
induce reference disagreement on correct transcripts by silencing a span's audio---overwriting every
sample over the target span's aligned interval---and sampling the model to see if it still produces the span.
Here $r$ is the silenced reference span and $a$ is any faithful alternative with the silenced span skipped.
\emph{Masked} \textsc{accept-ref} is the fraction of masked spans the model still emits---reproducing a span whose
audio is gone.
A `benchmaxxed' model would output the benchmark-specific reference content even after the relevant acoustic evidence has been removed.

Target spans can be selected by a number of criteria (e.g.\ random, prefix, suffix, etc.).
We target numbers because they are often hard to guess from the language model prior alone.
Surrounding audio qualities (pitch, environment, speaker) should provide little, if any, information
about a target number.
A faithful transcriber that cannot hear a number should rarely produce the exact reference value, so
recovery above the language-model prior ($\log p_M\!\bigl(r\mid x_\emptyset\bigr)$) indicates that the model has narrowed the reference distribution using benchmark-specific cues outside the masked span.

\subsection{Orthographic switching}
\label{sec:method-ortho}

Many tokens admit two renderings that are phonetically and semantically identical:
\emph{anyone} vs.\ \emph{any one}, \emph{Mr} vs.\ \emph{Mister}. We take $r$ to be the spelling the
local benchmark's reference used and $a$ the other spelling. Unlike the previous two behavioral probes, the audio
here supports both renderings equally, so a raw \textsc{accept-ref} is less informative. A model with
a fixed internal spelling (e.g. always output ``mister'') scores high whenever the corpus happens to share its bias. Thus, we also report on
``\emph{switch rate}'' $s$---the smaller of the two conditional \textsc{accept-ref} values.
A fixed-form transcriber scores $s=0.0$ and a model which randomly alternates
between the variants scores $s\le0.5$, so
$s$ reliably above $0.5$ signals that the model is
reading surrounding acoustic cues to match a given clip's convention. 

We test two convention
pairs: the \emph{honorific} (VoxPopuli always abbreviates \emph{Mr}; LibriSpeech spells \emph{Mister}),
which requires knowing which dataset a sample came from, and \emph{archaic spacing}
(\emph{any one}/\emph{anyone} etc.), both of whose forms occur within LibriSpeech itself, so switching
requires tracking sub-populations or individual samples rather than corpus-level conventions.
This probe complements the other two, extending evidence that models discriminate across datasets
and even naturally ambiguous cases \emph{within} a dataset.

\subsection{Mechanism localization}
\label{sec:method-localization}

The reference disagreement probe in particular
demonstrates that the model learns incorrect mappings from audio to text on the benchmark.
The implication is that these models have learned erroneous representations of the audio, and thus
might see degraded performance on real-world audio.
Alternatively, if models are able to localize the behavior to the specific acoustic cue of the benchmark, 
then they can faithfully transcribe the audio in other contexts, but apply the (erroneous) benchmark-optimized policy to 
still hill-climb on the benchmark. In this sense, the model would have inflated its reported performance on the benchmark
while neither destroying nor improving its real-world transcription ability.

\paragraph{What triggers the behavior.}
We characterize the trigger with a battery of contrast conditions, each holding the transcript fixed
while removing or replacing one component of the benchmark signal: the original recordings under additive
noise and reverberation; text-to-speech (TTS) clones of speakers from the dataset evaluation set; fresh speakers from the same domain (\textsc{ep-fresh} and \textsc{libri-fresh}); and control voices (generic TTS presets or DaiKon speakers).

We also truncate the audio to a short window around the target span; an audio-true transcription of
the isolated span shows the model can perceive it, in which case edits on the full clip are gated by the
surrounding context rather than by a perceptual failure.
In the reverse direction, we test whether the benchmark
signal alone re-ignites the behavior: an $8$\,s window of real test audio is appended to the clone conditions which previously did not trigger the benchmark-optimized behavior,
against a duration-matched conversational control donor.

The splice readout is a
difference-in-differences: the reported contrast is the flip rate under the benchmark donor minus
the flip rate under the control donor, so base difficulty and generic suffix effects cancel.
Each condition is read using the behavioral probe metrics, with the full reference disagreement \textsc{accept-ref} as
the primary readout and masked-number recovery as a secondary readout.

\paragraph{Where the audio is overridden.}
If a model can activate the benchmark-optimized policy to override faithful representations of generic audio simply by splicing
in benchmark audio, it suggests the model dedicates portions of the activation space to unpacking a benchmark's acoustic cue
and uses it to activate the benchmark-optimized behavior across the sequence.
To investigate further where the model activates the benchmark-optimized behavior,
we apply linear steering and activation patching to the model.

For activation patching, we use cases in which truncation led the model to transcribe the
audio-true rendering $a$. We replace the audio frames at the target span with the corresponding context-free encoding of the
audio-true span, with the assumption that those patched frames contain a faithful representation of $a$. If patching restores the audio-true transcript, the override was already written into the
encoder's representation; if the output does not change, the decoder would thus ignore even a faithful representation
of the span (likely attending to the rest of the audio encoding for the benchmark's audio cues). 
The activation patching
helps differentiate whether the encoder's representation of the target span's frames is the
issue compared to the decoder's policy with respect to the span.

Decoder-side edits can be further replicated
in select cases by asking models to output a direct translation of the audio (en$\to$es).
If a span omitted from the transcript is recovered in the translation
(App.~Table~\ref{tab:trunc}), it shows that the faithful representation is available, but is ignored by
the decoder's default transcription policy.
A similar result is found using an attention mask by testing if the
model retrieves an omitted span when forced
to attend to just the audio encoding frames of the target span (App.~Table~\ref{tab:trunc}).

For the linear steering probe, we learn a diff-in-means direction between suffix-spliced and control
inputs from $22$ \textsc{ep-fresh} courtesy splice pairs (App.~\ref{app:steer}), applied at a single
encoder layer.
The splice-control pair isolates how benchmark audio context (the spliced suffix) changes the encodings 
of earlier frames and therefore could approximate the benchmark-optimized encoder policy. 
If the policy is compact, adding the learned direction
should induce the benchmark-optimized behavior on audio where it previously generated an audio-true transcription. Projecting the vector, meanwhile, should
restore the audio-true transcription on real benchmark clips. 
This probe helps understand where in the network
the benchmark-specific routing is occurring and reinforces the theory that models are learning to selectively activate a benchmark-optimized policy.

Linear steering, truncation, and activation patching run on the general population of consensus-flagged edits. The
task-switch and attention levers instead require a fixed, position-anchored target span, and we
instantiate them on a case study: VoxPopuli's references systematically omit the audible opening courtesy
(``thank you, Mr President''), an omission the high-\textsc{accept-ref} models systematically reproduce.
\FloatBarrier
\section{Results}
\label{sec:results}

\begin{figure}[tbp]
\centering
\begin{subfigure}{0.45\textwidth}
  \centering\includegraphics[width=\linewidth]{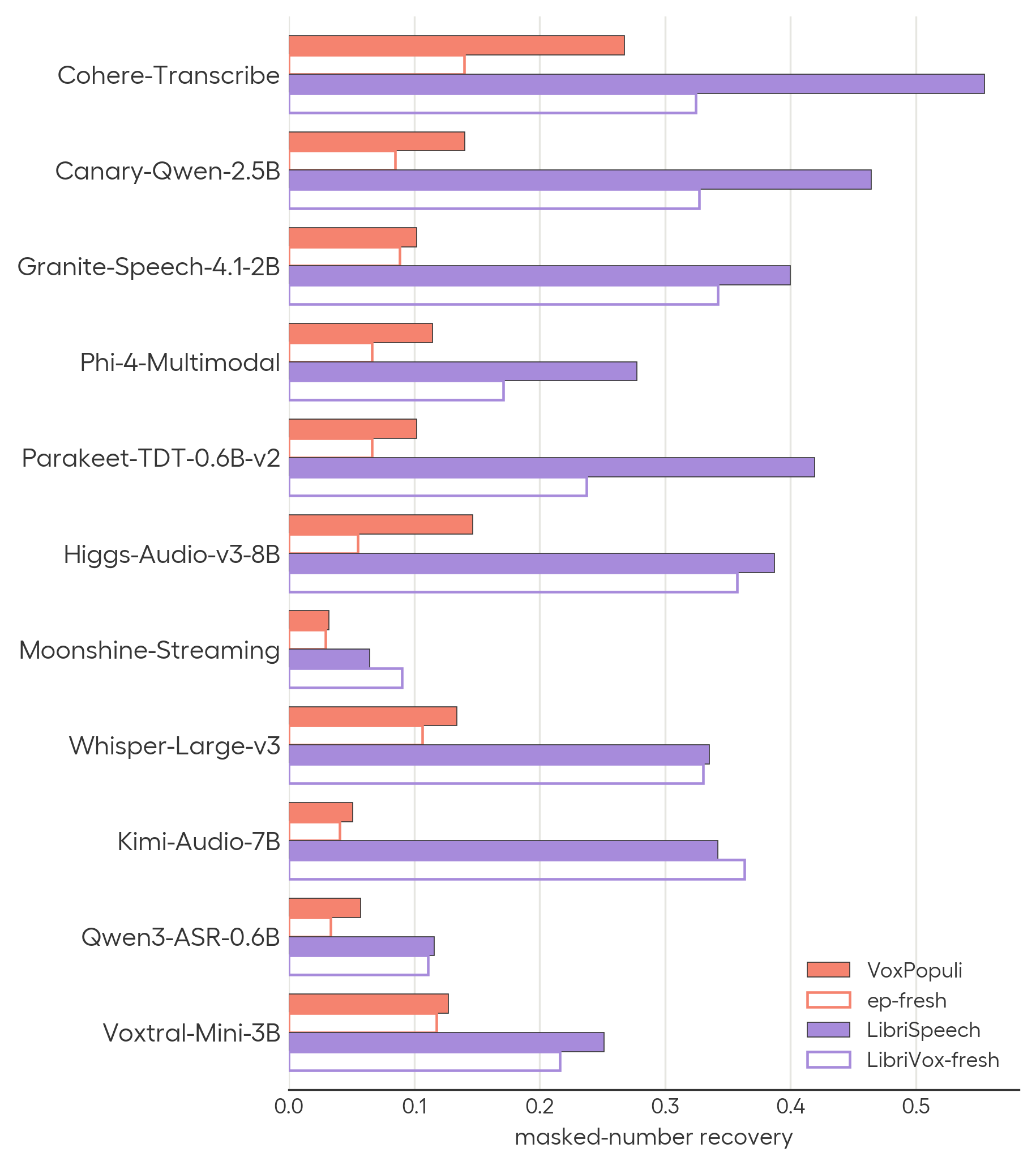}
  \caption{}\label{fig:mask-bb}
\end{subfigure}\hfill
\begin{subfigure}{0.45\textwidth}
  \centering\includegraphics[width=\linewidth]{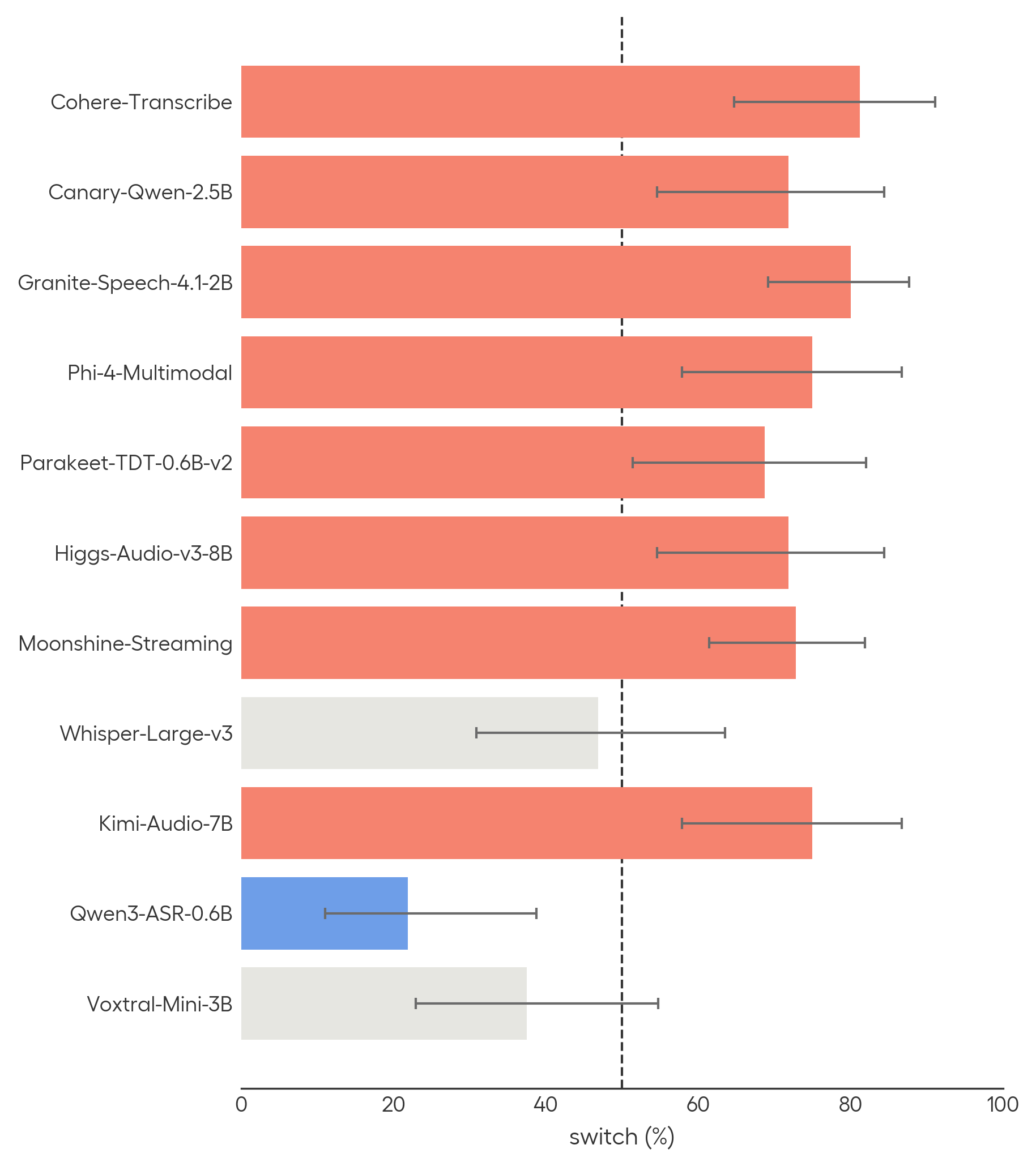}
  \caption{}\label{fig:ortho-spacing}
\end{subfigure}
\caption{(\subref{fig:mask-bb}) Masked-number \textsc{accept-ref} per corpus.
(\subref{fig:ortho-spacing}) Orthographic switch rate on archaic spacing.}
\label{fig:behavior-panels}
\end{figure}

For the reference disagreement probe, we find that
the six models with the best VoxPopuli WER ($5.4$--$5.8\%$)
are exactly those with the highest \textsc{accept-ref} ($0.18$--$0.30$), while every model at
$6.5\%$ WER or above sits at or below $0.10$.
The results are corroborated in terms of audio-lift per App.~Figure~\ref{fig:consensus-wb} and on human annotated data in App. ~Table~\ref{tab:aa-sensitivity}.
\begin{figure}[tbp]
\centering
\includegraphics[width=0.66\textwidth]{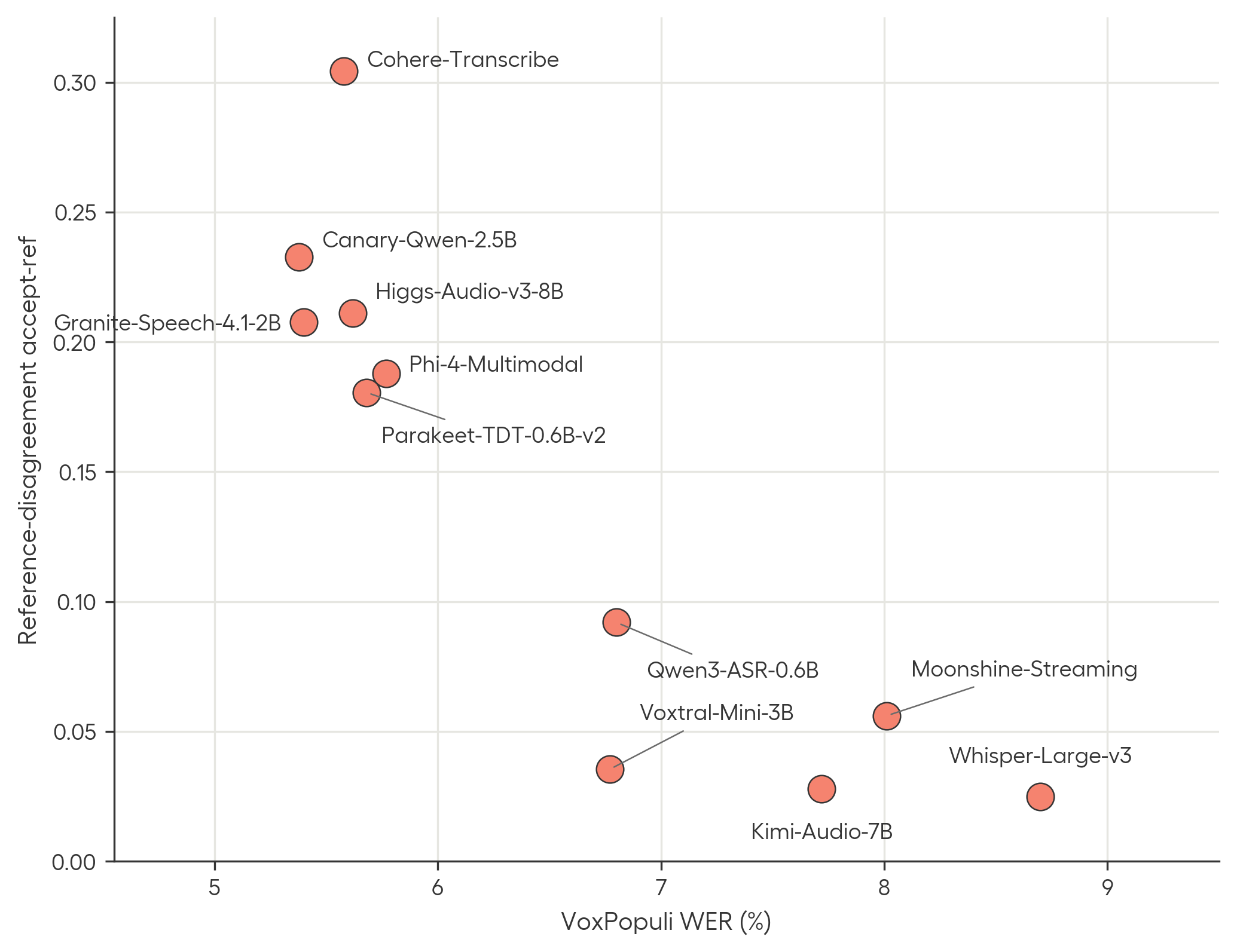}
\caption{Cross-model audit on VoxPopuli. WER (\%) is the VoxPopuli-test
score from the June 2026 Open ASR Leaderboard~\citep{srivastav2025openasrleaderboardreproducible}.
Kimi Audio is not on the leaderboard, and its score is computed using the leaderboard's
scoring.
Consensus-panel members are scored against edits flagged unanimously by the remaining
three members (\S\ref{sec:method-consensus}).}
\label{fig:consensus}
\end{figure}
Several of the same models had slightly higher masked \textsc{accept-ref}  on the public benchmarks (VoxPopuli, LibriSpeech---where the top
models reach ${\sim}0.40$) than on held-out sets \textsc{libri-fresh} and \textsc{ep-fresh},
a finding reinforced by the audio-lift numbers (Figure~\ref{fig:behavior-panels}\subref{fig:mask-bb} and App. ~Table~\ref{tab:nummask-lift}).
On the honorific switch, 
six out of 11 models significantly exceed the 0.5 baseline switch rate. 
On archaic spacing (Figure~\ref{fig:ortho-spacing}), eight of 11 models exceed
$0.5$ switch rate, suggesting that models are able to discriminate not just between datasets but also
between subpopulations or even individual samples of
the same dataset. We observed no significant regression in \textsc{accept-ref} for non-leaked speakers in VoxPopuli.

\FloatBarrier
\subsection{A narrow acoustic context gates the behavior}
\label{sec:results-analysis}

Voice clones of speakers from the VoxPopuli and LibriSpeech evaluation sets trigger directionally similar \textsc{accept-ref} to the original audio. However, when a generic voice reads the identical transcript,
\textsc{accept-ref} falls for many models.

Clones of the fresh, same domain speakers often rest closer to the generic \textsc{accept-ref},
which suggests the trigger for the
benchmark-optimized policy is narrowly attached to the benchmark-specific cue and does not 
generalize to new data from a similar domain. These findings are clearest for reference-disagreement \textsc{accept-ref} on VoxPopuli, but the effect survives the language-model-prior control: on the masked probe, \textsc{audio lift} is positive on the real recordings and on clones of dataset speakers but collapses toward zero on generic voices for the elevated models (Cohere, Canary-Qwen, Phi-4, Higgs; App.~Fig.~\ref{fig:trig-lift}).
A positive audio lift could potentially be explained by the fact that models can attend to the entire audio sequence,
which contains semantic representations of the entire sequence that are not present in a pure language
causal mask. However, if audio lift falls for the same sentence when using a generic voice, it shows that the model is not simply attending to lookahead semantics
to unmask the number, but rather depending on the specific acoustic cue of the benchmark audio to more easily
decode the number.
\begin{figure}[tbp]
\centering
\includegraphics[width=0.42\textwidth]{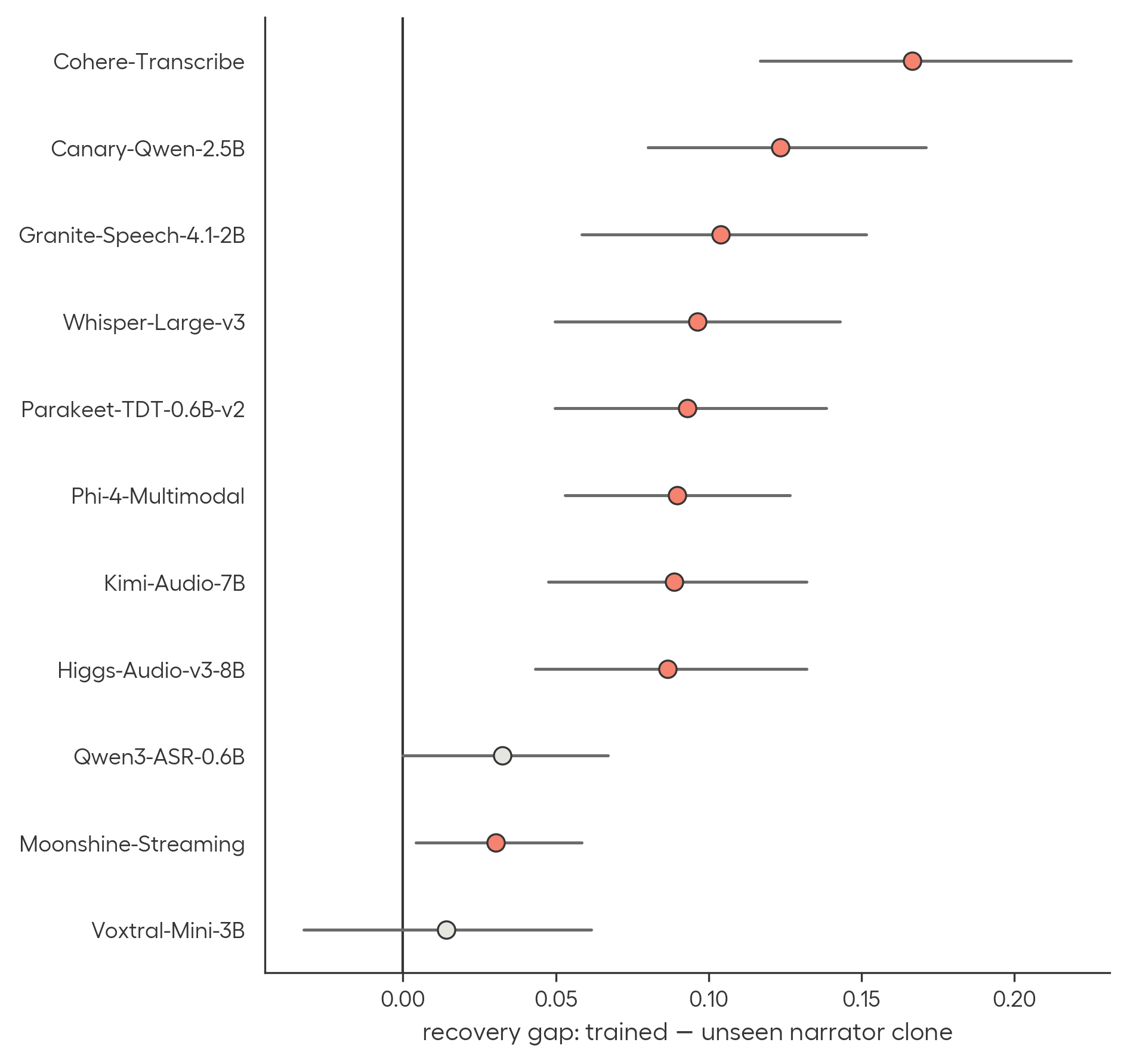}
\caption{Difference in masked-number recovery between LibriSpeech test-set narrator clones and held-out \textsc{libri-fresh} narrator clones reading identical sentences (test-set minus held-out; positive values indicate greater recovery for test-set narrator clones). Sentence-clustered bootstrap 95\% confidence intervals.}
\label{fig:voice-gap}
\end{figure}

\begin{figure}[tbp]
\centering
\begin{subfigure}{0.49\textwidth}
  \centering\includegraphics[width=\linewidth]{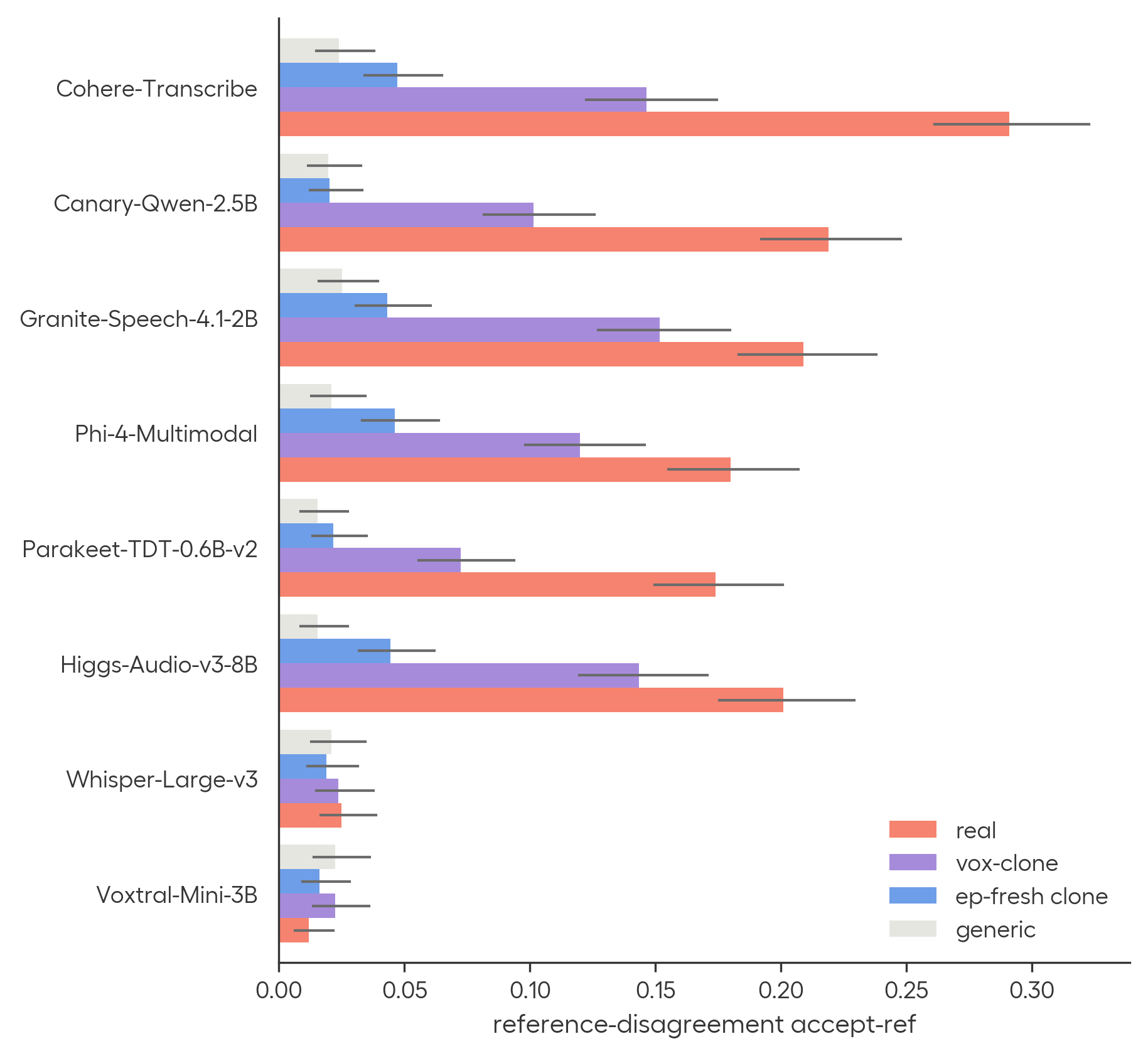}
  \caption{}\label{fig:consensus-tightness}
\end{subfigure}\hfill
\begin{subfigure}{0.49\textwidth}
  \centering\includegraphics[width=\linewidth]{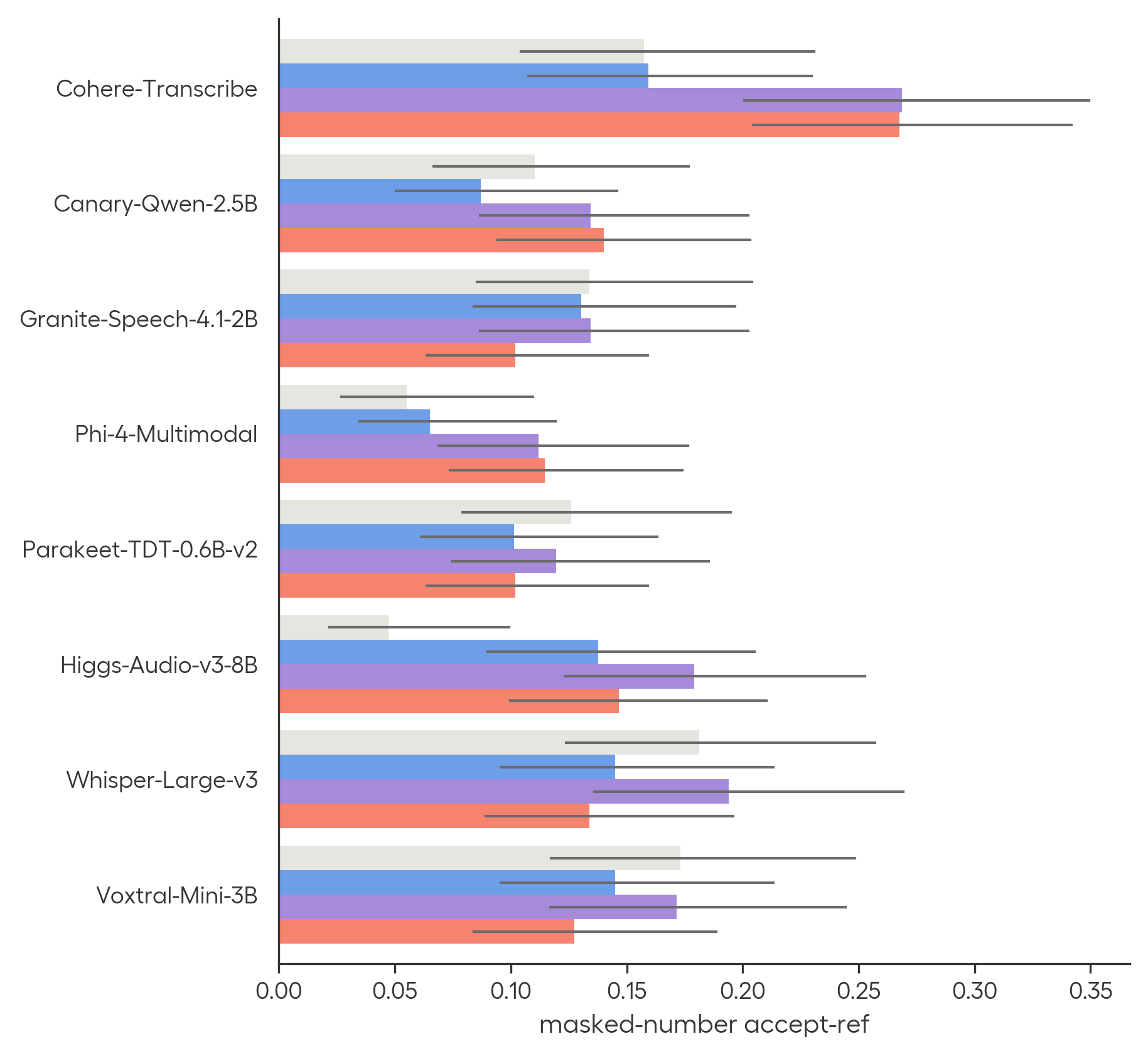}
  \caption{}\label{fig:battery-masked}
\end{subfigure}\\[2pt]
\begin{subfigure}{0.49\textwidth}
  \centering\includegraphics[width=\linewidth]{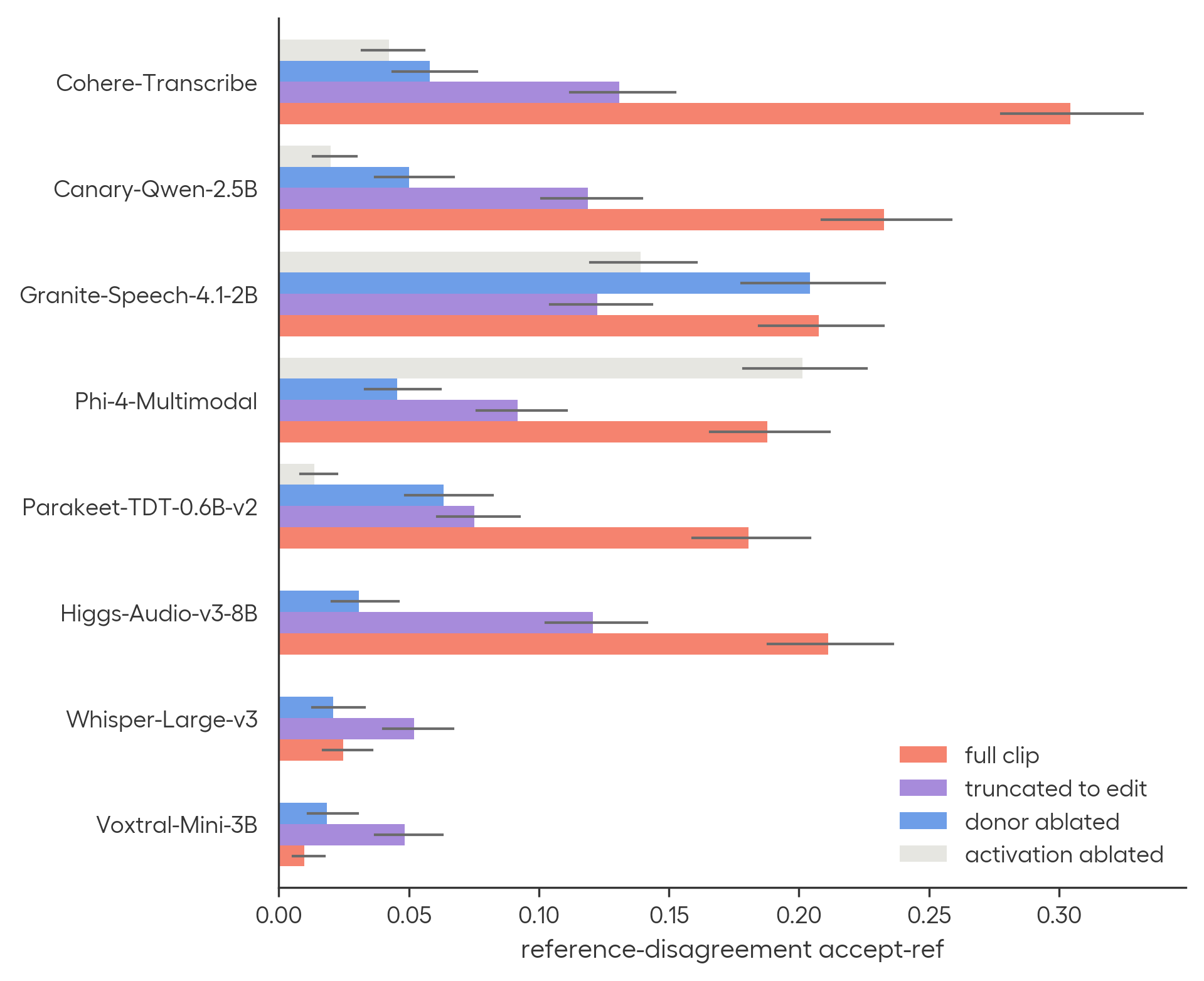}
  \caption{}\label{fig:battery-ablation}
\end{subfigure}\hfill
\begin{subfigure}{0.49\textwidth}
  \centering\includegraphics[width=\linewidth]{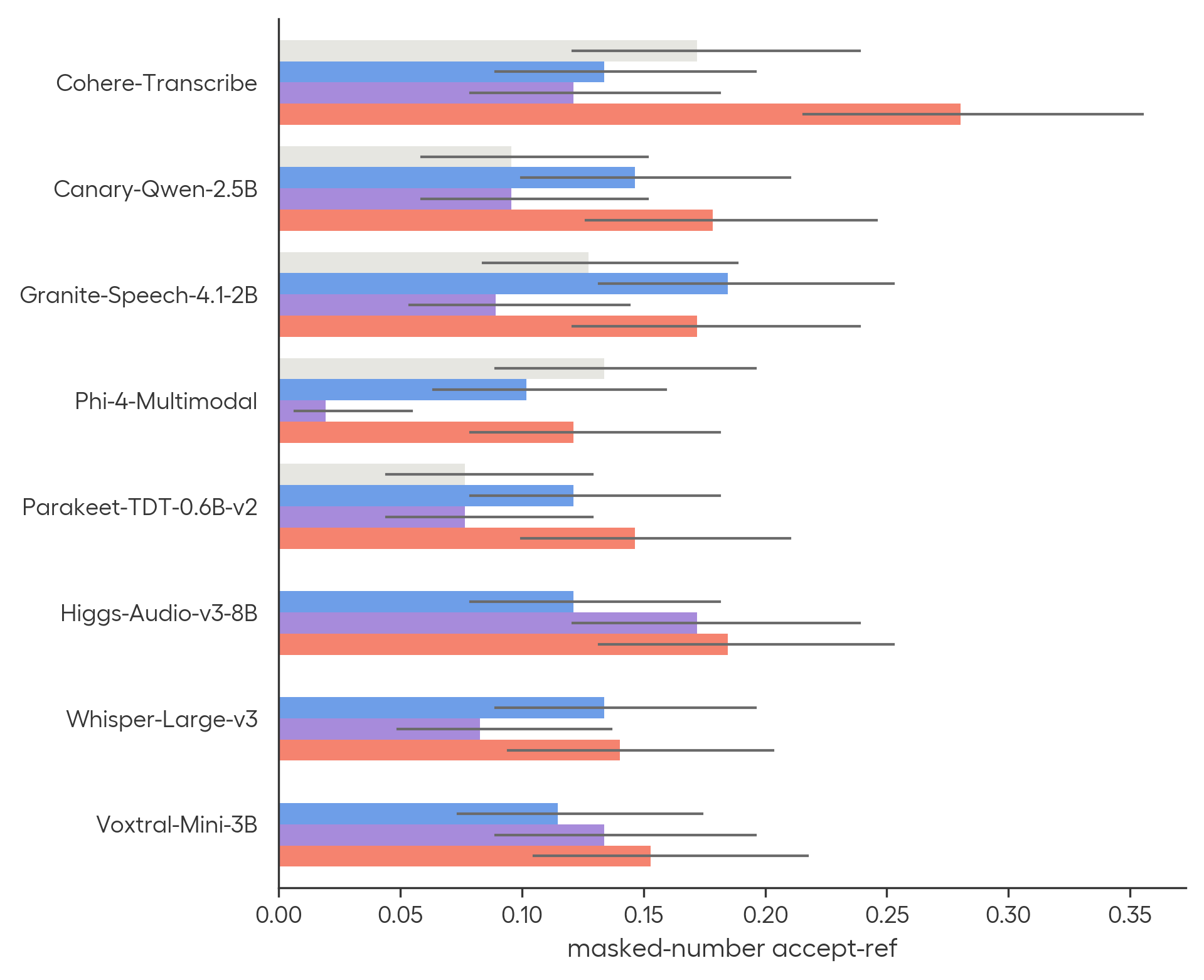}
  \caption{}\label{fig:battery-masked-ablation}
\end{subfigure}
\caption{The trigger battery, VoxPopuli. Top row: voice conditions on identical
transcripts---(\subref{fig:consensus-tightness}) reference-disagreement and
(\subref{fig:battery-masked}) masked-number \textsc{accept-ref}. Bottom row: the same probes with the
trigger removed instead of the voice varied---truncation to the edit, an appended $8$\,s conversational
donor, or the learned register direction projected out of one encoder layer
(\subref{fig:battery-ablation}, \subref{fig:battery-masked-ablation}). Wilson $95\%$ CIs.}
\label{fig:trigger-battery}
\end{figure}

Meanwhile, truncating the audio to a short segment
around the target span collapses
reference-disagreement and masked \textsc{accept-ref} toward the floor for several models
(Figure~\ref{fig:trigger-battery}\subref{fig:battery-ablation}). The more the surrounding context marks the audio as coming from the
benchmark, the more the model reproduces its reference errors; diluting those cues removes the
behavior. 

\begin{figure}[tbp]
\centering
\begin{subfigure}{0.92\textwidth}
  \centering\includegraphics[width=\linewidth]{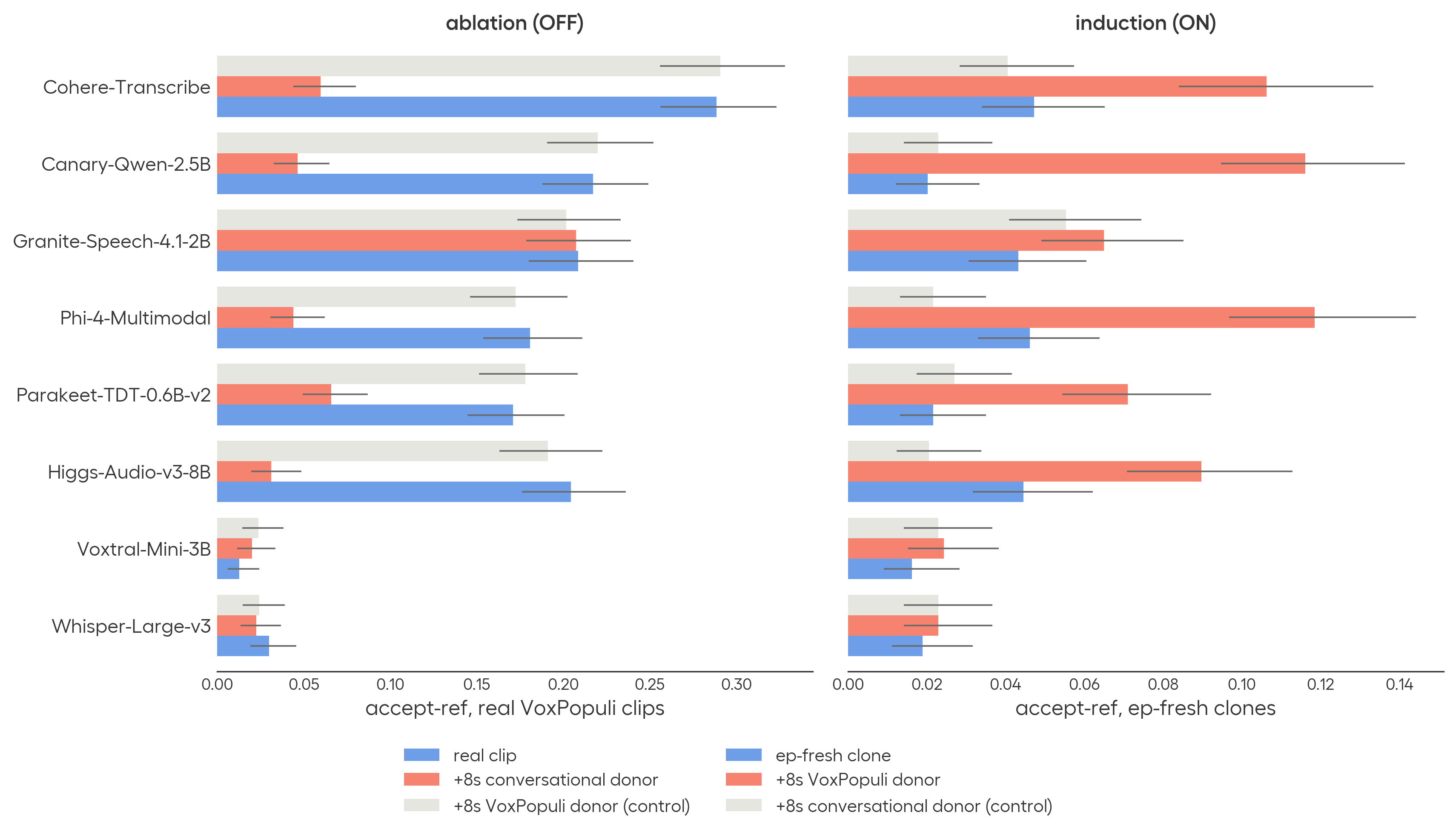}
  \caption{}\label{fig:steer-input}
\end{subfigure}\\[4pt]
\begin{subfigure}{0.92\textwidth}
  \centering\includegraphics[width=\linewidth]{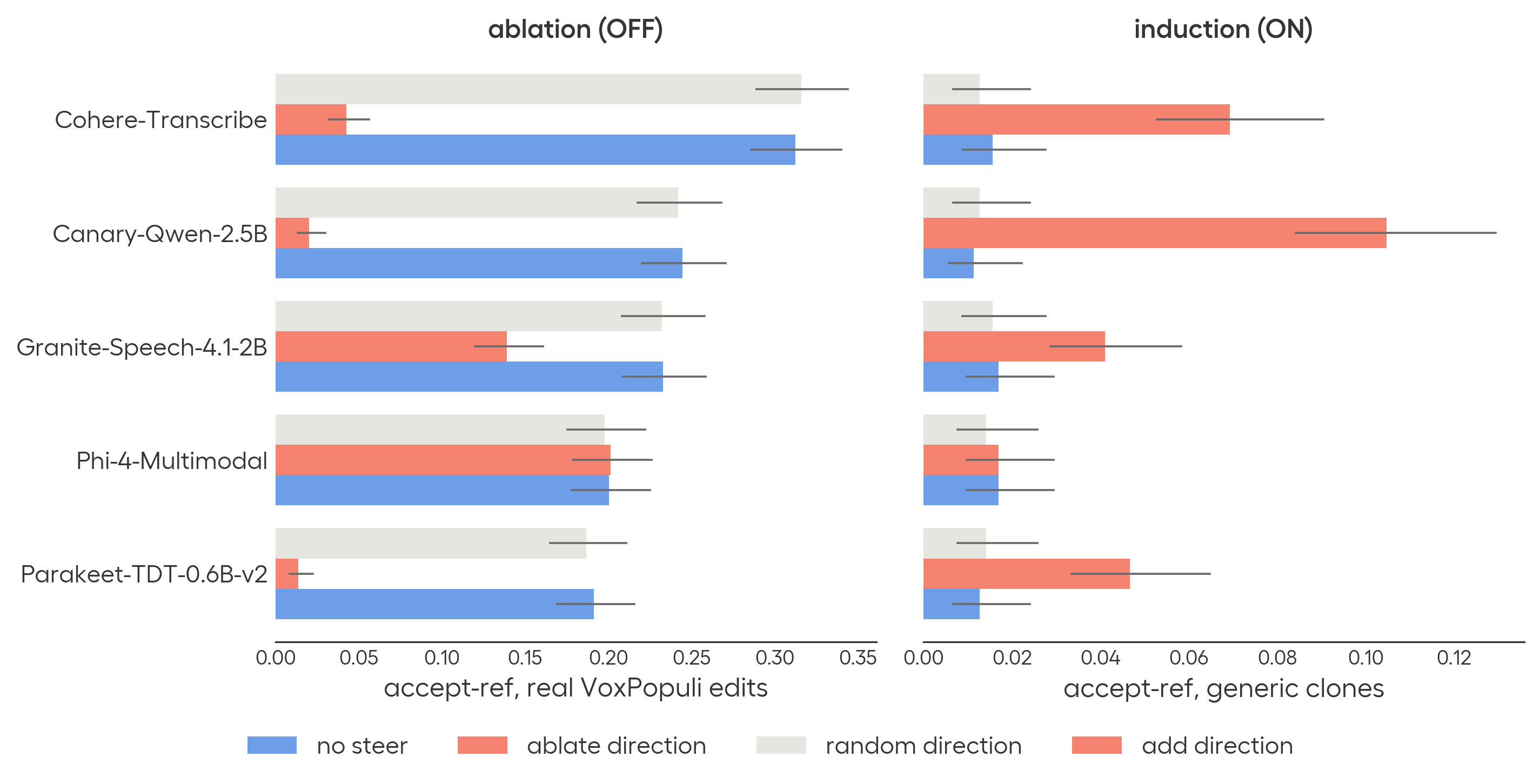}
  \caption{}\label{fig:steer-act}
\end{subfigure}
\caption{Switching the benchmark-optimized policy on and off. (\subref{fig:steer-input}) Input level: On real clips (top left) a
conversational donor collapses \textsc{accept-ref} while a VoxPopuli donor
leaves it intact; on \textsc{ep-fresh} clones of the same sentences (top right) a VoxPopuli donor
re-ignites it while the conversational donor does not.
(\subref{fig:steer-act}) Activation level: projecting out the learned direction on real benchmark edits
(bottom left) and adding it on generic-voice clones (bottom right).
The remaining consensus-panel members show no effect, like Voxtral-Mini-3B.}
\label{fig:steer-causal}
\end{figure}

Appending a generic donor from the conversational dataset to the real
benchmark clips collapses \textsc{accept-ref} for Cohere, Canary-Qwen, Phi-4,
Parakeet, and Higgs, while an appended VoxPopuli donor leaves it unchanged
(Figure~\ref{fig:steer-causal}\subref{fig:steer-input}).
Likewise, appending a ``donor'' VoxPopuli
clip to synthetic \textsc{ep-fresh} with low \textsc{accept-ref} bumps the \textsc{accept-ref} (\textsc{ep-fresh}-clone bases: Phi-4 $+.10$, Canary
$+.09$, Higgs $+.07$, Cohere $+.07$, Parakeet $+.04$ over a duration-matched control donor).

Testing the models under acoustic perturbations to the original audio further suggests that models rely on different subsets of benchmark-associated cues: additive noise removes the behavior for some models, while others retain it under both noise and reverberation (App.~Figure~\ref{fig:trig-robustness}).

In sum, when the models transcribe
reference transcripts with contradictory audio, models don't learn a general
incorrect mapping  (e.g. they don't lose the ability to transcribe
a literal phrase such as ``thank you, Mr. President'' in all circumstances). 
When the spans in question are said in voices outside the distribution of the benchmark's
evaluation set, the model outputs an audio-true transcription. 
Even audio that normally would trigger the benchmark-optimized policy reverts to the
audio-true transcription by adding non-benchmark audio to the clip or by removing enough benchmark context around the span. 
In this sense, the trigger of the benchmark-optimized policy is relatively `narrow.'
The models with the highest accept-ref rate in particular are able to determine fairly precisely whether an audio clip is in
the VoxPopuli dataset, generalizing across speakers in the test set, but not to new speakers from a more recent parliamentary recording.

\subsection{The behavior spans encoder and decoder and is causally steerable}
From our activation patching probes, we find evidence of both the encoder and decoder suppressing faithful representations, showing that the model learns the benchmark-optimized behavior end-to-end
as opposed to simply via natural language data contamination~\citep{tseng2025evaluation}. 
For reference
insertions, replacing the target-frame encoding with its context-restricted counterpart often
restored the audio-supported output, consistent with the edit being represented in the encoder
state. For deletions and substitutions, the same intervention was less effective, suggesting that
later decoding behavior can override a locally faithful representation
(App.~Figure~\ref{fig:patch-dissociation}).
 Attention restriction and translation provided additional evidence that faithful information
sometimes remained available despite being omitted during transcription (App.~Table~\ref{tab:trunc}).

The learned direction bidirectionally steers four of the six elevated
models---Cohere, Parakeet, Canary, and (partially) Granite---with a low-rank structure for the first
three ($k{=}1$ recovers $65$--$80\%$ of the full-direction effect).
Adding the direction to held-out generic audio (which previously had near-zero accept-ref rates)
mildly raises the model's overall \textsc{accept-ref} rate
($0.02{\to}0.07$ Cohere, $0.01{\to}0.05$ Parakeet, $0.01{\to}0.11$ Canary, and $0.02{\to}0.04$ Granite
over the same $707$ gated edits).
Ablating the same direction has a stronger effect:
 \textsc{accept-ref} falls by $82$--$92\%$ when the steering vector is projected onto Cohere, Canary, and Parakeet.

These results show that benchmark-conditioned transcription behavior can be causally modified at both
the input and activation levels in a subset of models. 

\FloatBarrier
\section{Discussion and Conclusion}
\label{sec:discussion}
We introduced a behavioral methodology---reference-error reproduction, masked-entity recovery, and orthographic
convention switching---for measuring ASR benchmark optimization in cases where the audio
underdetermines the reference. Across state-of-the-art open models on two widely used benchmarks, reproducing benchmark conventions
against the audio is common among the highest-scoring systems. We also employ a mechanistic methodology using synthetic audio, task switching, activation patching,
and attention readouts to understand when and how these models
activate this benchmark-optimized behavior. We found that models are able to activate benchmark-optimized
behavior based on narrow acoustic cues, to the point where
we can bidirectionally flip the benchmark-optimized behavior by appending audio or steering a single model layer.

Using surrounding acoustic context is not itself undesirable: prosody, speaker characteristics, and recording context can legitimately affect transcription. Our probes instead target cases where those cues do not justify a preference for the benchmark reference over an acoustically supported alternative.
Our results tentatively suggest that benchmark optimization is a particularly relevant concern for high-dimensional modalities like audio.
Unlike traditional autoregressive LMs, ASR models generally can attend to the entire audio encoding,
which contains not only semantic content but also speaker, channel, and other acoustic information across the sequence.
This gives models additional degrees of freedom for learning benchmark-specific policies based on narrow acoustic cues,
even on relatively short, underdetermined sequences.
As reinforcement learning (RL) becomes increasingly prevalent for speech language models \citep{shivakumar2025group}, understanding these dataset-specific acoustic cues
becomes even more important and warrants further study as a potential source of reward hacking.

We offer three practical takeaways for the field from our research.

First, for model developers, model releases should be as transparent as possible around their training data.
We investigated how these behaviors operate at inference time, but do not yet fully understand how they arise during training.
Having more transparent data mixtures in general would help the field further study whether these behaviors arise from benchmark-guided model selection algorithms, data leakage, memorization, or other mechanisms.
For instance, based on the available public information about training recipes, the models with the highest \textsc{accept-ref} (Phi-4, Cohere-Transcribe, Granite, and Canary) trained on less data than
the models with low \textsc{accept-ref} (e.g., Qwen3, Whisper). Indeed, Qwen3 does display mild benchmark-optimized inclinations, such as occasionally
firing the courtesy trigger, yet this behavior is rare compared to other models. Qwen trained on 40 million hours of weakly supervised data,
whereas most of the high-\textsc{accept-ref} models trained on less than 1 million hours of data. Further analysis here could help the field
better understand the scale of data needed to avoid these behaviors and how they arise during training.
Moreover, some of the behavioral probes and mechanistic probes from this paper can be used as sanity checks during and after training
to regularize models.

Second, for benchmark development, we recommend against using i.i.d. test splits for benchmark evaluation.
As the lack of generalization of benchmark optimization behaviors to the {\textsc{ep-fresh}} dataset shows,
temporal or metadata (e.g., speaker) stratification is minimally necessary to understand the generalization of model behavior.
The publicly available VoxPopuli test dataset on Hugging Face mixes $40\%$ of speakers in the test set into the training distribution.
It should be noted, however, that no model showed a leaked-speaker advantage on our probes (pooled
elevated-model reference-disagreement \textsc{accept-ref} $0.22$ on leaked vs.\ $0.26$ on unleaked
speakers; masked recovery likewise indistinguishable), which suggests that official train-test speaker overlap alone does not explain the observed behavior.
More ideally, the main test sets that model providers report performance on should be fully held-out, non-public evaluation sets so they can neither
be leaked nor gamed as easily \citep{ayllon2026rw, bezzam2026benchmaxxer}.

Third, for practitioners selecting ASR models, we recommend looking at multiple metrics besides just WER on public benchmarks.
This is especially true of VoxPopuli, which has a high rate of reference errors, to the point that any model below
$3\%$ WER has to transcribe reference errors. Our work uses a consensus reference edit procedure which
aligned with human annotations and can be scaled to help improve the usefulness of VoxPopuli as a dataset.
Moreover, our mechanistic and behavioral probes provide additional insight into a model's behavior beyond WER 
to better understand how optimized a model is for a given benchmark.
\FloatBarrier

\bibliographystyle{plainnat}
\bibliography{refs}\FloatBarrier
\newpage
\appendix
\section{Methodological details}
\label{app:methods}

\subsection{Text processing and teacher-forced NLL}
\label{app:localization}
All transcripts are scored under the standard Hugging Face Open ASR leaderboard text normalizer from June 2026 (lowercasing, punctuation
removal, number and contraction normalization). Negative Log Likelihood (NLL) quantities are computed by teacher-forcing the relevant reference
string and summing per-token log-probabilities over the target span; for encoder--decoder
models we force the decoder on the reference with the audio encoder conditioned on the clip, and for speech--LLM models we force the text continuation after the
audio prefix.

The numerator of the white-box readout (Eq.~\ref{eq:lift}) is the \emph{unnormalized span
log-likelihood ratio}, obtained by summing the per-token log-likelihood differences over the query span.
This sum is invariant to subword segmentation, unlike a per-token mean. Dividing it by the number of
characters in the reference span gives $\lambda(r)$ and makes spans of different lengths commensurate.
Across clips we
report the corpus mean with a $2000$-resample percentile-bootstrap $95\%$ CI.

\subsection{EOS masking}
\label{app:eos}

At each scored position we remove end of sequence tokens (EOS) from the softmax denominator before taking the log-probability,
renormalizing the distribution over continuation tokens; we apply this symmetrically to the audio term and
the $x_\emptyset$ prior term, as masking only one would compare two differently-normalized distributions.
We mask EOS because a subset of models place substantial probability on EOS when the
language-model prior is extracted from fully silenced audio without the standard system prompt:
 Moonshine ($0.76$) and Granite ($0.61$) most strongly, Qwen3-ASR moderately
($0.25$). The remaining models place $\approx\!0$ probability on the EOS token.
The choice is also the
conservative one for our claims---it suppresses the measured audio lift, since taking unnormalized
logits would shrink the decoder prior while leaving the audio-conditioned scores nearly unchanged.

\subsection{Synthetic speech stimuli and the intelligibility gate}
\label{app:ttsgate}

All synthetic samples are
generated with Qwen3-TTS.  Generic
uses one of the stock voices from Qwen3 1.7B CustomVoice model. The cloned voices use the Qwen3 Base model with a speaker from the relevant dataset; the Vox-cloned case uses a reference clip from a
VoxPopuli evaluation-set speaker who is not the speaker in the original clip.

Because a TTS rollout can garble the intended sentence, every synthetic clip passes an intelligibility
gate before use: at least one of the eleven models must transcribe the entire intended transcript
exactly (zero WER under the harness normalizer). Pass rates are $0.84$--$0.93$ across the gated sets
(e.g.\ $568/679$ on the consensus-flagged generic renderings). 

\subsection{Alignment, masking, and number selection}
\label{app:masking}

\paragraph{Forced alignment.} We obtain per-word time spans with the Massively Multilingual Speech
forced-alignment (\textsc{mms-fa}) model~\citep{pratap2024mms}.
\textsc{mms-fa} emits connectionist temporal classification (CTC)
frame indices, which we convert to seconds with the per-clip ratio
$\texttt{(n\_samples} / \texttt{n\_frames}) / 16000$. Reference words are lowercased and stripped and numeric tokens are expanded to their
spoken form via the \texttt{num2words} package.

Words that normalize to empty such as pure punctuation are assigned
a zero-width span at the preceding word boundary so that the emitted span list stays index-aligned ($1{:}1$)
with the whitespace-split reference. We
align only clips of at least $4$\,s to remove tiny segments when the mask would span a significant percentage of the audio.

\paragraph{Masking a span.} Given a target word's aligned $[t_0,t_1]$, the masking procedure silences just that span. We mask every occurrence of the target value in the clip so the scored token cannot leak from a repeated mention elsewhere.
 We pad the mask by $\texttt{mask\_pad}=120$\,ms to more conservatively handle imperfections in the aligner.
 In cases when the aligner's boundaries are suspiciously tight ($t_1-t_0<60$\,ms), we instead mask the entire inter-word gap,
and skip the clip when that gap is itself $<80$\,ms (about $20\%$ of number targets). These drops bias the sample
toward longer, harder numbers, exactly the cases in which language priors should not be as helpful.
We exclude the number ``one'' because it is hard to distinguish from its variants
(e.g. ``one must not'') and is often trivial to predict.

\paragraph{Hard-cell difficulty gate.} The headline masked readouts are reported ungated (the audio
lift handles the prior by subtracting it). As a check that black-box recovery is not language-prior
guessing, we re-score the VoxPopuli masked column on \emph{hard cells}: spans whose silenced-audio
prior NLL/char, $\mathrm{nll}_\emptyset(r) = -\log p_M(r \mid x_\emptyset)/|r|_{\mathrm{char}}$, taken as the median over the
white-box-instrumented models, is at least $3.5$ nats---$62$ of the $92$ covered spans ($67\%$). The
same hard subset scores every model; a per-model gate would select a different subset per model and the
rates would not be comparable. On hard cells recovery \emph{rises} for the elevated models (Cohere
$0.27{\to}0.36$, Canary $0.14{\to}0.24$, Higgs $0.15{\to}0.23$, Granite $0.10{\to}0.19$) and stays at
the floor for the near-zero models (Moonshine $0.00$, Kimi and Qwen3 $0.07$); Phi-4 is flat ($0.11$).
Recovery thus concentrates on exactly the spans the prior cannot supply.

\subsection{Content-preserving perturbations and the robustness protocol}
\label{app:perturb}

To test whether the reference-following behavior survives the acoustic degradations used to build
derived benchmarks (\S\ref{sec:related}), we re-run the accept-reference and masked-number probes on
perturbed copies of the real audio. All perturbations are content-preserving digital signal processing (DSP) applied to the
$16$\,kHz mono waveform (implementation in \texttt{scripts/vmt/perturb.py}); each is available as a severity
ladder, of which we report a single moderate level per family:
\begin{itemize}
\item \textbf{Additive noise}---Gaussian noise scaled to a target signal-to-noise ratio (SNR) $\in\{20,10,5,0\}$\,dB (reported: $10$\,dB).
\item \textbf{Reverberation}---convolution with a measured room impulse response \cite{ko2017rirs}, binned by $\mathrm{RT}_{60}$ into
$[0.2,0.4]$, $[0.45,0.7]$, $[0.85,1.3]$\,s and reported at the middle bin (median
$\mathrm{RT}_{60}$ $0.52$\,s). 

Both are
truncated to the input length so the perturbed clip is sample-for-sample duration-matched
to the source, then energy-matched to the original RMS.

\end{itemize}
These families mirror the corruptions of recent augmented benchmarks
\citep{shah2025srb,xu2026megaasr,goswami2026whisperrirmega}.

\paragraph{Interpretation.} Additive noise at $10$\,dB has little effect on overall WER, so changes in
\textsc{accept-ref} are unlikely to result from a broad loss of transcription ability. Strong reverberation
raises WER for some models, making changes in \textsc{accept-ref} harder to interpret. Reverberation can
also carry some of a masked word's audio beyond the masked interval, so masked-number results under
reverberation provide weaker evidence than those under noise.

\subsection{Splice-induction stimuli}
\label{app:splice}

A stimulus is $[\,\text{base}\,][\,0.15\,\mathrm{s\ silence}\,][\,\text{donor window}\,]$ where a donor can be placed as a suffix or prefix to the base clip. Donor windows are cut mid-utterance from
VoxPopuli-test and root-mean-square (RMS)-matched to the base clip's loudness. The control donor is a duration-matched window of
conversational audiobook speech.

Base clips are the clone renderings of the consensus-flagged transcripts (generic-voice and
\textsc{ep-fresh}-speaker; \S\ref{sec:results-analysis}) where models correctly transcribed the clip rather than transcribing the reference errors and, for the steering training pairs, real
\textsc{ep-fresh} recordings (App.~\ref{app:steer}).

\subsection{Activation steering protocol}
\label{app:steer}

The training stimuli are the input-level splice pairs of \S\ref{sec:results-analysis}: $43$
\textsc{ep-fresh} courtesy-opener clips (held-out post-cutoff parliamentary content so no behavioral flip enters the
selection), each rendered as $[\text{base}][0.15\,\mathrm{s\ silence}][5\,\mathrm{s\ donor}]$ with a real
VoxPopuli donor vs.\ a duration-matched audiobook donor. The first
$22$ bases train the direction; the remaining $21$ are held out.
For each model we hook every encoder layer and mean-pool activations over the base-clip frames
only. The direction is therefore the suffix's effect on the representation of the
identical base content---carried into the base frames by the encoder's bidirectional context
mixing. The steering direction at
layer $L$ is the diff-in-means $d_L = \mathrm{mean}(\text{vox-suffix}) - \mathrm{mean}(\text{control-suffix})$
over the $22$ training pairs, unit-normalized.
Induction adds $\alpha\,d_L$ at a single layer during decoding; ablation projects the direction out.
Evaluation is on held-out clips disjoint from the training pairs, with a garbage gate discarding
degenerate outputs (which leads to dropping most outputs for Higgs), and norm-matched random directions as the
specificity floor. Rank structure is assessed by restricting $d_L$ to the top-$k$ components from a principal component analysis (PCA) of the
training diffs; $k{=}1$ recovers $65$--$80\%$ and $k{=}4$ $85$--$100\%$ of the full-direction induction
rate for Cohere ($k{=}1$: $0.71$), Parakeet ($0.76$), and Canary ($0.67$). Best single-layer ablation
sites: Cohere L$37$ $\alpha{=}4$, Parakeet L$12$ $\alpha{=}4$, Canary L$12$ $\alpha{=}2$, Granite L$14$
$\alpha{=}4$, Phi-4 L$17$ $\alpha{=}4$. Granite's effect is partial in both directions (ablation flips
$33\%$ of reference reproductions; induction $0.017{\to}0.041$), consistent with its decoder re-imposing
part of the behavior. Phi-4's direction separates the spliced conditions but is causally inert---ablation
$0.200{\to}0.201$ and induction are flat---consistent with Phi's policy being driven by the decoder parameters over the encoder parameters. Higgs-Audio's direction (L$11$, held-out area under the receiver operating characteristic curve (AUC) $0.82$) has no
clean operating point: at $\alpha{=}4$ ablation degrades decoding wholesale (WER against the
consensus-corrected transcript $0.06{\to}0.27$, $36$ garbage clips) while at $\alpha{\le}2$ it is inert
($0.186{\to}0.174$)---so we report no activation cell for Higgs. The same input-level donor lever demonstrably moves both models
(Figure~\ref{fig:steer-causal}\subref{fig:steer-input}; Phi-4 donor ablation $0.19{\to}0.05$ and the
largest \textsc{ep-fresh} splice induction, Higgs $0.20{\to}0.03$)---the failure is in reproducing that
switch with a single-layer linear edit of the encoder, not in the ability to steer these models in general.

The generalized-ablation readout decodes three arms per clip in one pass---no steer, ablate, and
random---on the panel-unanimous consensus edit on VoxPopuli test ($745$ clips; $1{,}113$ edits). Per edit, the verdict (reproduces the erroneous
reference vs.\ follows the audio) is recomputed on each arm's fresh hypothesis; flips count only when both arms are non-garbage. Ablation flips $88\%$ (Cohere, $303$ flips), $95\%$ (Parakeet, $201$),
$93\%$ (Canary, $252$), and $45\%$ (Granite, $115$) of no-steer reference reproductions to the audio-true
rendering, with $\le11$ regressions per model; the random arm is flat. WER against the original (erroneous) reference rises
while WER against the consensus-corrected transcript falls $25$--$40\%$.
The direction also transfers across probes for Cohere: on the real masked-number clips of the trigger
battery, ablating it halves recovery of the silenced number ($0.102{\to}0.051$, $n{=}157$; $10$ lost
vs.\ $2$ gained paired flips, exact McNemar $p{=}.039$) while the random arm is exactly flat. Canary and
Parakeet move in the same direction at small counts ($5{\to}3$ and $6{\to}4$ recoveries); Granite and
Phi-4 do not move.

The induction readout mirrors this on generic-voice clones of the consensus-corrected
transcripts ($495$ gated clips; $707$ unanimous edits), decoding four arms per clip: no steer, add
$\alpha\,d_L$ at $\alpha{=}4$ and $\alpha{=}8$, and a norm-matched random direction at $\alpha{=}8$.
Adding the direction raises \textsc{accept-ref} from $0.016$ to $0.072$ (Cohere; already at
$\alpha{=}4$), $0.013$ to $0.047$ (Parakeet), $0.011$ to $0.105$ (Canary), and $0.017$ to $0.041$
(Granite), with no garbage outputs and the random arm flat ($0.011$--$0.015$); the lift appears in all three edit classes (e.g.\ Cohere
insertions $0.03{\to}0.19$, deletions $0.02{\to}0.06$, substitutions $0.01{\to}0.06$). A masked-number
readout on generic masked clones is directionally positive at $\alpha{=}4$ but small and inconsistent
at $n{=}147$.

\subsection{Supplementary results}
\label{app:supp}

This section collects supporting figures and tables referenced from the main text.

\paragraph{Trigger sufficiency conditions.} The splice conditions of the trigger battery
(\S\ref{sec:method-localization}) append an $8$\,s mid-utterance window of a distinct real
VoxPopuli-test donor speaker (RMS-matched, $0.15$\,s gap) to each base clip, against a duration-matched
conversational control donor; $8$\,s is approximately the median VoxPopuli test utterance duration.
Bases are the generic-voice and \textsc{ep-fresh}-speaker clone renderings of the consensus-flagged transcripts. 
Reference-disagreement vox$-$control contrasts (clip-level bootstrap $95\%$ CIs): on ep-fresh-clone bases Phi-4 $+.096$
$[.074,.120]$, Canary $+.093$ $[.073,.115]$, Higgs $+.069$ $[.050,.091]$, Cohere $+.066$ $[.046,.087]$,
Parakeet $+.044$ $[.027,.062]$, Granite $+.010$ (not significant); on generic bases Phi-4 $+.018$ $[.006,.030]$, Cohere
$+.015$ $[.003,.027]$; every other model is within noise of zero in both settings.

\begin{figure}[tbp]\centering
\includegraphics[width=0.6\textwidth]{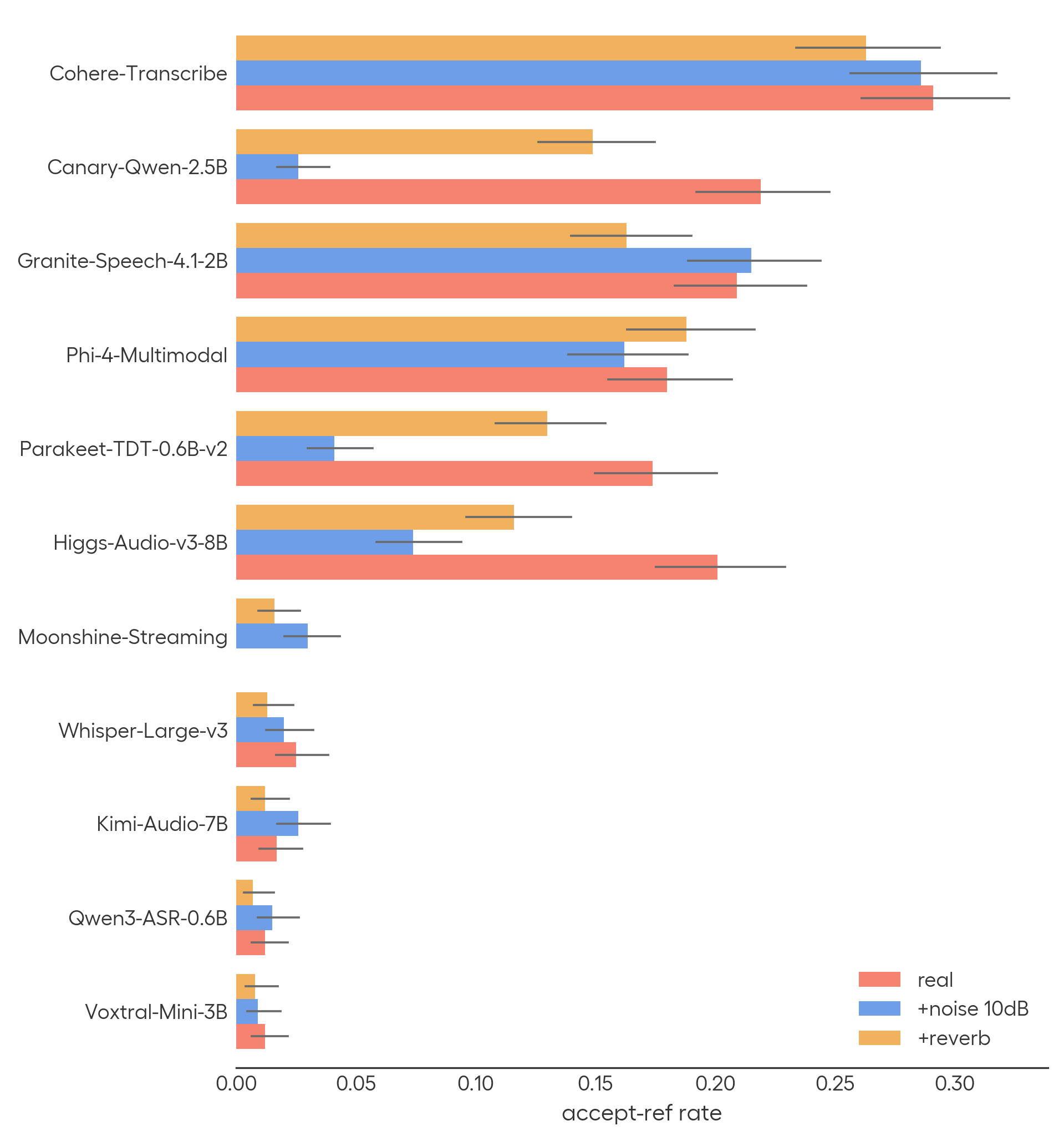}
\caption{Reference-disagreement \textsc{accept-ref} on the real VoxPopuli recordings under
content-preserving perturbations (additive noise $10$\,dB; measured room reverberation, RT60 $0.60$).
Wilson $95\%$ CIs.}
\label{fig:trig-robustness}
\end{figure}

\paragraph{Perturbation conditions.} Corrupting the real recordings while holding voice and content
fixed dissociates the models (Figure~\ref{fig:trig-robustness}): additive noise ($10$\,dB) alone
collapses Canary-Qwen, Parakeet, and Higgs to near zero ($.22\!\to\!.03$,
$.17\!\to\!.04$, $.20\!\to\!.07$), while Cohere, Granite, and Phi-4 retain the behavior under noise and measured
room reverberation (Cohere $.29\!\to\!.29/.26$).

\begin{table}[tbp]\centering\small
\caption{\textsc{accept-ref} on VoxPopuli-AA human-annotated edits, beside the consensus rates of
Figure~\ref{fig:consensus}. Ordering is nearly identical aside from Granite and Canary swapping places.}
\label{tab:aa-sensitivity}

\begin{tabular}{l cc c c}
\toprule
& consensus & \multicolumn{3}{c}{human-annotated edits} \\
\cmidrule(lr){2-2}\cmidrule(lr){3-5}
model & \textsc{accept-ref} & \textsc{accept-ref} & $95\%$ CI & $n$ \\
\midrule
Cohere-Transcribe                & 0.30 & 0.52 & [0.47, 0.58] & 253/483 \\
Granite-Speech-4.1-2B            & 0.21 & 0.42 & [0.36, 0.47] & 211/508 \\
Canary-Qwen-2.5B                 & 0.23 & 0.41 & [0.36, 0.47] & 210/507 \\
Higgs-Audio-v3-8B                & 0.21 & 0.39 & [0.33, 0.44] & 200/517 \\
Phi-4-Multimodal                 & 0.19 & 0.38 & [0.33, 0.44] & 195/510 \\
Parakeet-TDT-0.6B-v2             & 0.18 & 0.38 & [0.32, 0.43] & 192/512 \\
Qwen3-ASR-0.6B      & 0.09 & 0.19 & [0.15, 0.24] & 98/518 \\
Moonshine-Streaming & 0.06 & 0.14 & [0.11, 0.19] & 74/516 \\
Voxtral-Mini-3B     & 0.04 & 0.09 & [0.06, 0.12] & 47/527 \\
Kimi-Audio-7B       & 0.03 & 0.07 & [0.05, 0.11] & 40/536 \\
Whisper-Large-v3                 & 0.02 & 0.08 & [0.05, 0.11] & 39/514 \\
\bottomrule
\end{tabular}

\vspace{2pt}
\end{table}

\begin{figure}[tbp]
\centering
\includegraphics[width=\linewidth]{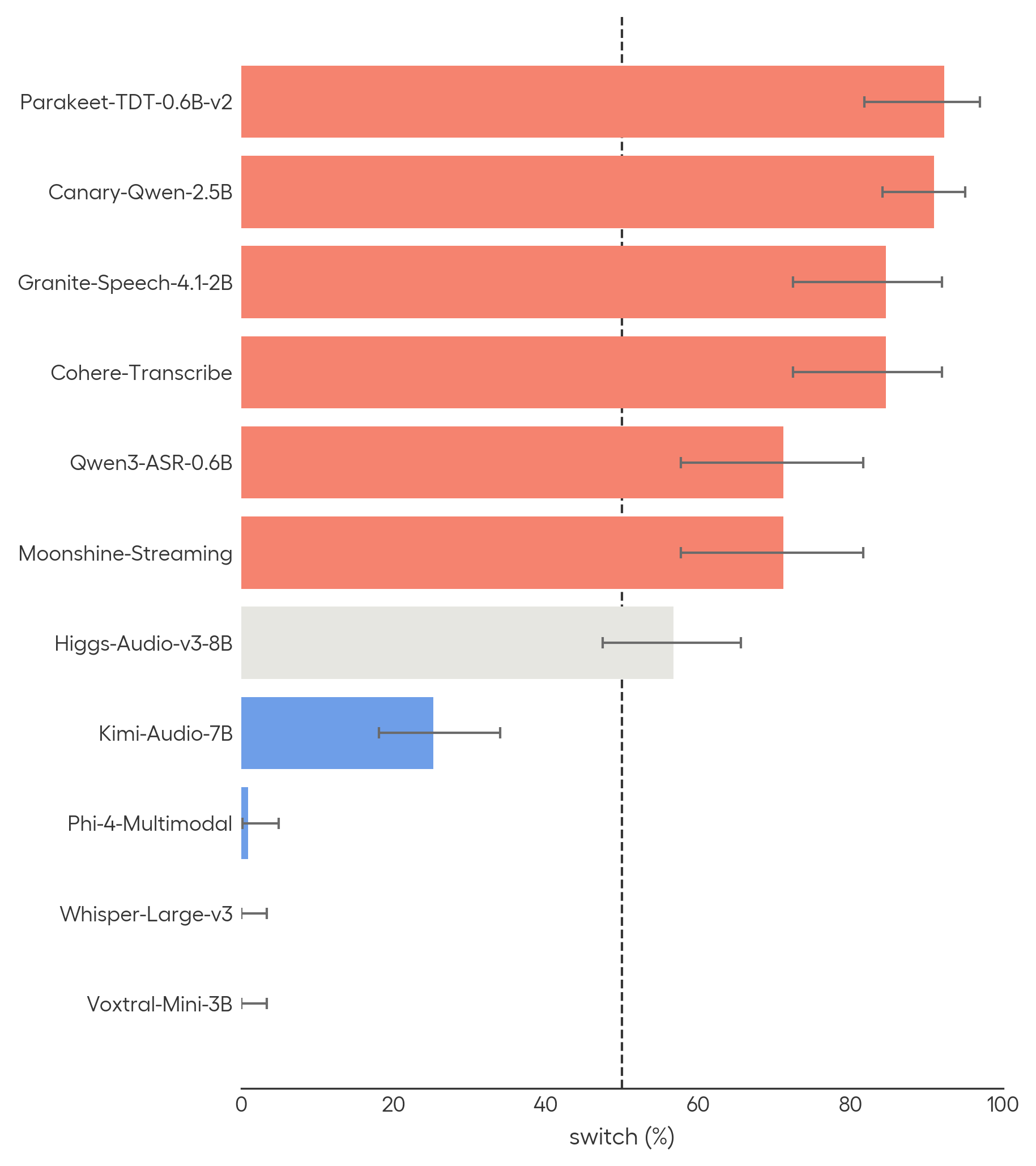}
\caption{Honorific switch rate (\emph{Mr}/\emph{Mister}), all 11 models. A rate above $0.5$ (dashed) means the model tracks
each corpus's convention at rates above chance.}
\label{fig:ortho-mister}
\end{figure}

\begin{figure}[tbp]
\centering
\includegraphics[width=\linewidth]{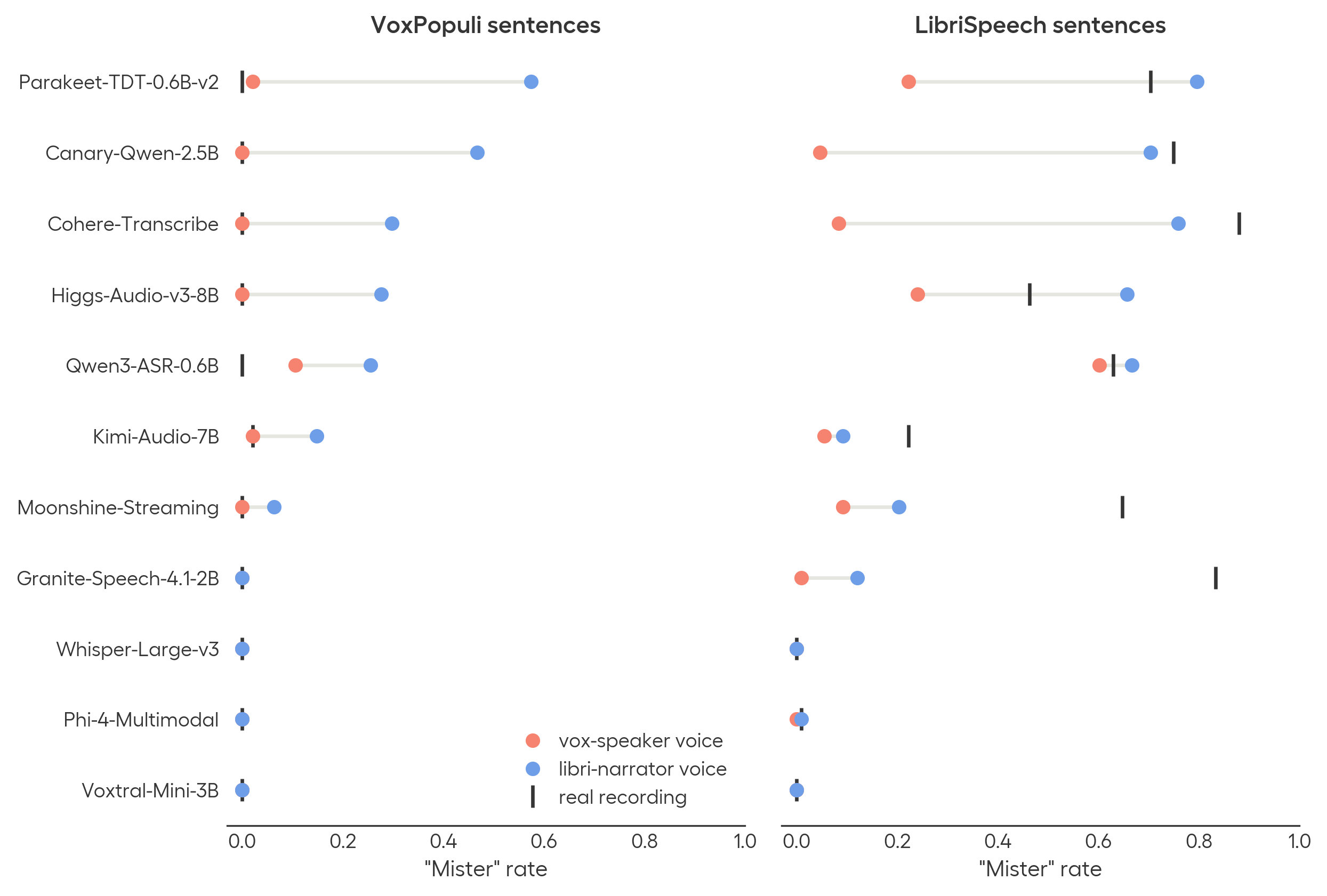}
\caption{Base-anchored \emph{Mister} rate for the same
sentence rendered in a cloned VoxPopuli-speaker voice vs.\ a cloned LibriSpeech-narrator voice; ticks
mark the real-recording rate. Left: VoxPopuli \emph{Mr} sentences ($n{=}47$; voice transfer). Right:
LibriSpeech \emph{Mister} sentences ($n{=}108$; voice erosion). Granite retains conventions on real
audio but loses them under any TTS clone, so its clone cells are relatively uninformative; Whisper,
Phi-4, and Voxtral never emit \emph{Mister}.}
\label{fig:hon-voice}
\end{figure}

\begin{figure}[tbp]
\centering
\begin{subfigure}{0.49\textwidth}
\centering\includegraphics[width=\linewidth]{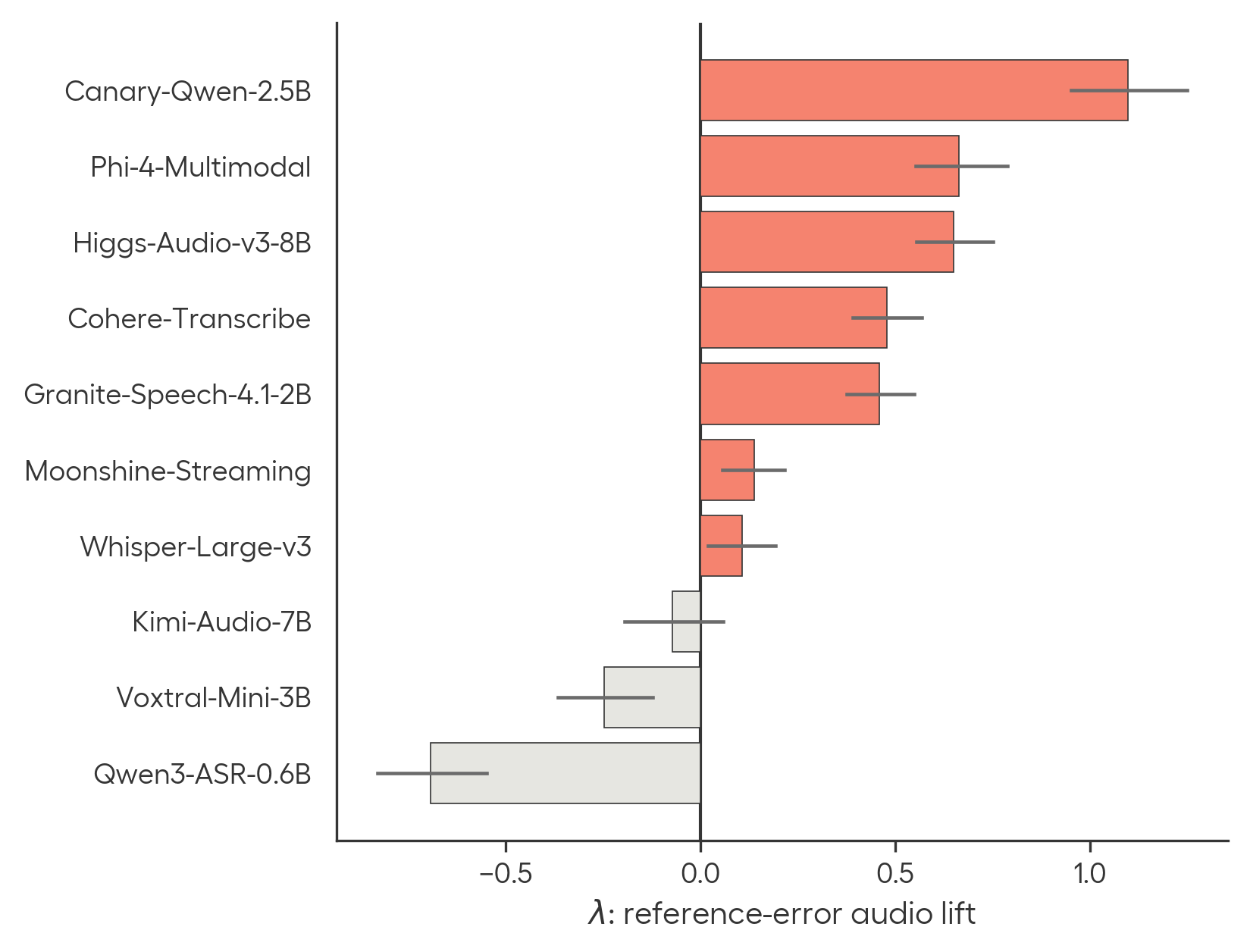}
\caption{}\label{fig:consensus-wb}
\end{subfigure}\hfill
\begin{subfigure}{0.49\textwidth}
\centering\includegraphics[width=\linewidth]{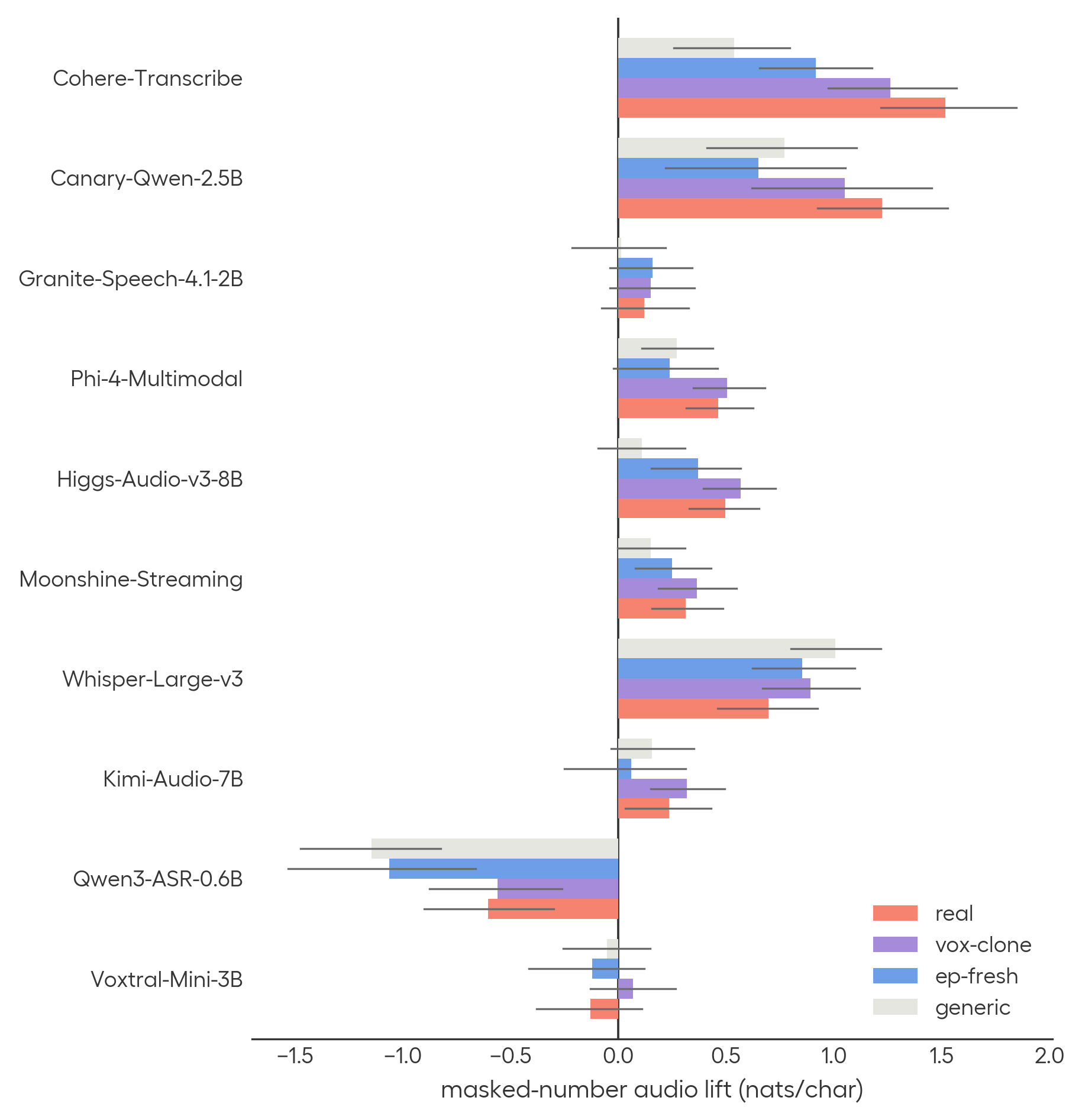}
\caption{}\label{fig:trig-lift}
\end{subfigure}
\caption{Audio lift $\lambda(r)$ (Eq.~\ref{eq:lift}), in nats per character; bootstrap $95\%$ CIs. Parakeet-TDT is excluded
(no teacher-forced readout). (\subref{fig:consensus-wb}) The reference disagreement audio lift. (\subref{fig:trig-lift}) The masked entity audio lift for different voice conditions for $115$ paired
sentences.}
\label{fig:audio-lift}
\end{figure}

\begin{table}[tbp]\centering\small
\caption{Reference-disagreement \textsc{accept-ref} on consensus edits as audio context is
removed or the trigger is ablated. \emph{truncated} cuts the audio to a tight window around the edit span ($\pm1$
aligned word $\pm0.25$\,s); \emph{donor ablated} appends an $8$\,s conversational donor
to the full clip; \emph{activation ablated} projects the learned register direction out of a single
encoder layer.}
\label{tab:isolate}
\begin{tabular}{l ccccc}
\toprule
model & full & truncated & donor ablated & activation ablated \\
\midrule
Cohere-Transcribe                  & 0.30 & 0.13 & 0.06 & 0.04 \\
Canary-Qwen-2.5B                   & 0.23  & 0.12 & 0.05 & 0.02 \\
Granite-Speech-4.1-2B              & 0.21  & 0.12 & 0.20 & 0.14 \\
Higgs-Audio-v3-8B                  & 0.21  & 0.12 & 0.03 & -- \\
Phi-4-Multimodal                   & 0.19  & 0.09 & 0.05 & 0.20 \\
Parakeet-TDT-0.6B-v2               & 0.18  & 0.08 & 0.06 & 0.01 \\
Qwen3-ASR-0.6B         & 0.09  & 0.09 & 0.04 & -- \\
Moonshine-Streaming    & 0.06 & 0.07 & 0.04 & -- \\
Voxtral-Mini-3B        & 0.04 & 0.05 & 0.03 & -- \\
Whisper-Large-v3                   & 0.02 & 0.05 & 0.02 & -- \\
Kimi-Audio-7B          & 0.03 & 0.06 & 0.04 & -- \\
\bottomrule
\end{tabular}

\end{table}
\begin{table}[tbp]\centering\small
\caption{Full-probe \textsc{accept-ref} on real audio with each corpus's own content: VoxPopuli-test vs
\textsc{ep-fresh}. Masked columns score the corpus-paired subsets, so the VoxPopuli masked rates differ
from the full masked set of \S\ref{sec:results}.}
\label{tab:realfresh}

\begin{tabular}{l cc cc}
\toprule
& \multicolumn{2}{c}{consensus \textsc{accept-ref}} & \multicolumn{2}{c}{masked \textsc{accept-ref}} \\
\cmidrule(lr){2-3}\cmidrule(lr){4-5}
model & VoxPopuli & ep-fresh & VoxPopuli & ep-fresh \\
\midrule
Cohere-Transcribe                  & 0.304 & 0.122 & 0.185 & 0.074 \\
Canary-Qwen-2.5B                   & 0.233 & 0.150 & 0.051 & 0.062 \\
Granite-Speech-4.1-2B              & 0.208 & 0.117 & 0.051 & 0.062 \\
Higgs-Audio-v3-8B                  & 0.211 & 0.168 & 0.070 & 0.040 \\
Phi-4-Multimodal                   & 0.188 & 0.193 & 0.076 & 0.044 \\
Parakeet-TDT-0.6B-v2               & 0.181 & 0.120 & 0.038 & 0.029 \\
Qwen3-ASR-0.6B         & 0.092 & 0.091 & 0.051 & 0.015 \\
Moonshine-Streaming    & 0.056 & 0.099 & 0.013 & 0.018 \\
Voxtral-Mini-3B        & 0.035 & 0.128 & 0.076 & 0.062 \\
Whisper-Large-v3                   & 0.025 & 0.071 & 0.083 & 0.062 \\
Kimi-Audio-7B          & 0.028 & 0.144 & 0.013 & 0.018 \\
\bottomrule
\end{tabular}
\end{table}

\begin{figure}[tbp]
\centering
\includegraphics[width=0.72\textwidth]{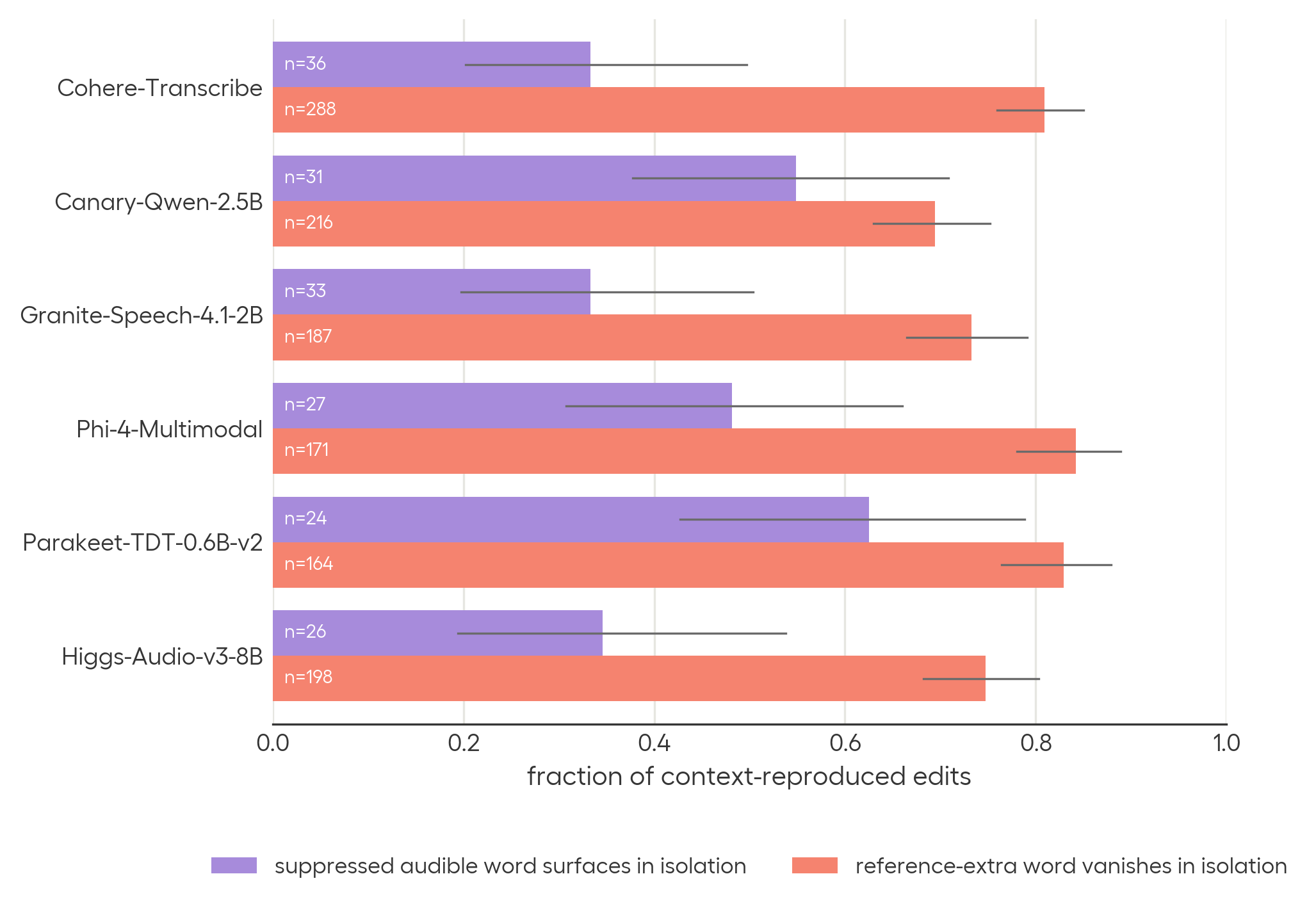}
\caption{Context gating on all consensus edits, conditioned on reproduction: for each model with a high \textsc{accept-ref}, the
edits it reproduces on the full clip, re-decoded on the isolated window of Table~\ref{tab:isolate}.
Reference-extra words mostly disappear and suppressed audible words resurface once the surrounding
benchmark context is cut away. Wilson $95\%$ CIs.}
\label{fig:isolate-gating}
\end{figure}

\begin{figure}[tbp]
\centering
\includegraphics[width=0.72\textwidth]{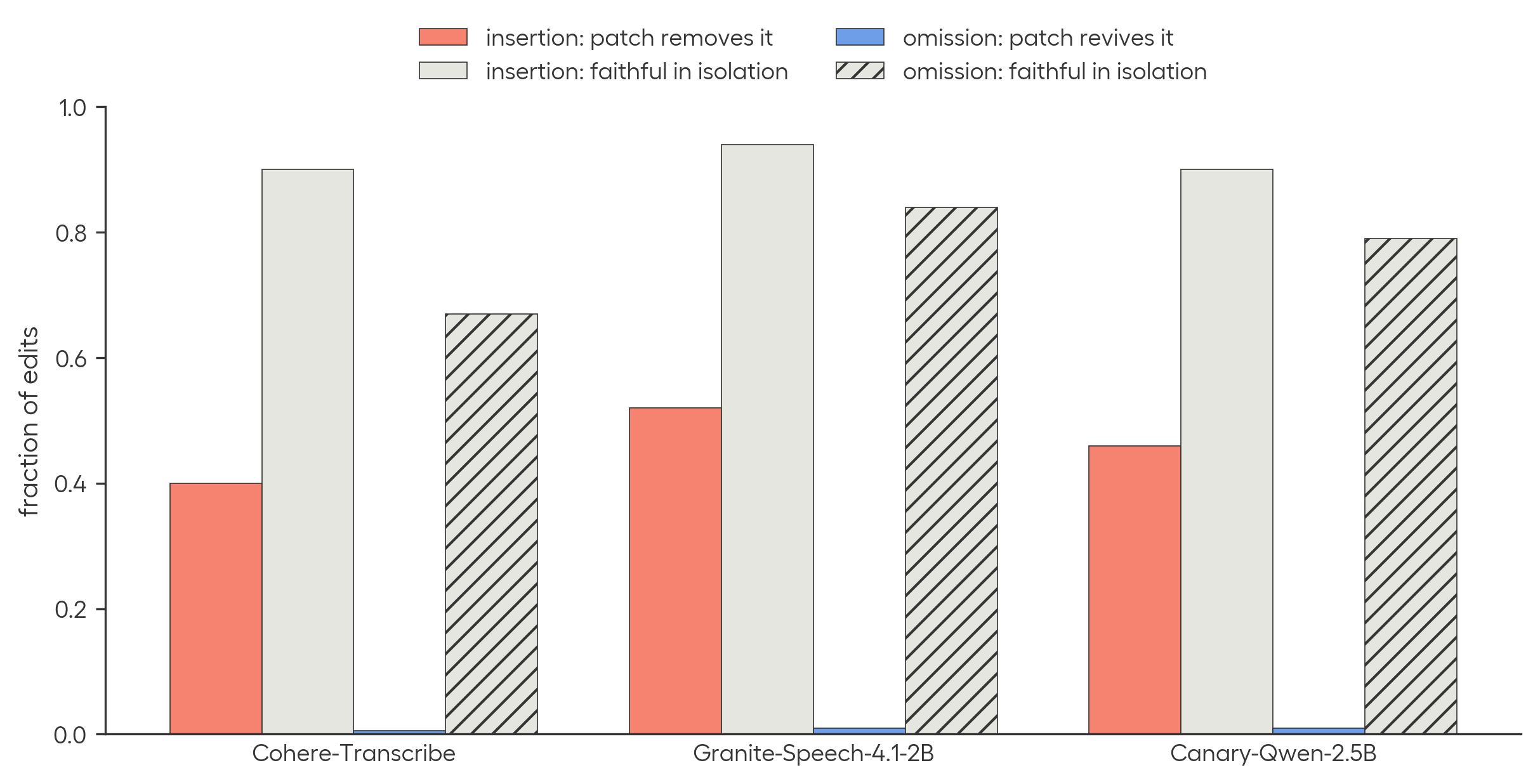}
\caption{Edit-locus dissociation (Cohere, Granite, Canary-Qwen). Isolating the span restores faithful
transcription for both edit types (context gating; the all-model version is
Figure~\ref{fig:isolate-gating}). Patching in a context-free encoding of the span removes reference
\emph{insertions} (encoder-side) but never restores reference \emph{omissions} (decoder-side).}
\label{fig:patch-dissociation}
\end{figure}

\begin{figure}[tbp]\centering
\includegraphics[width=0.6\textwidth]{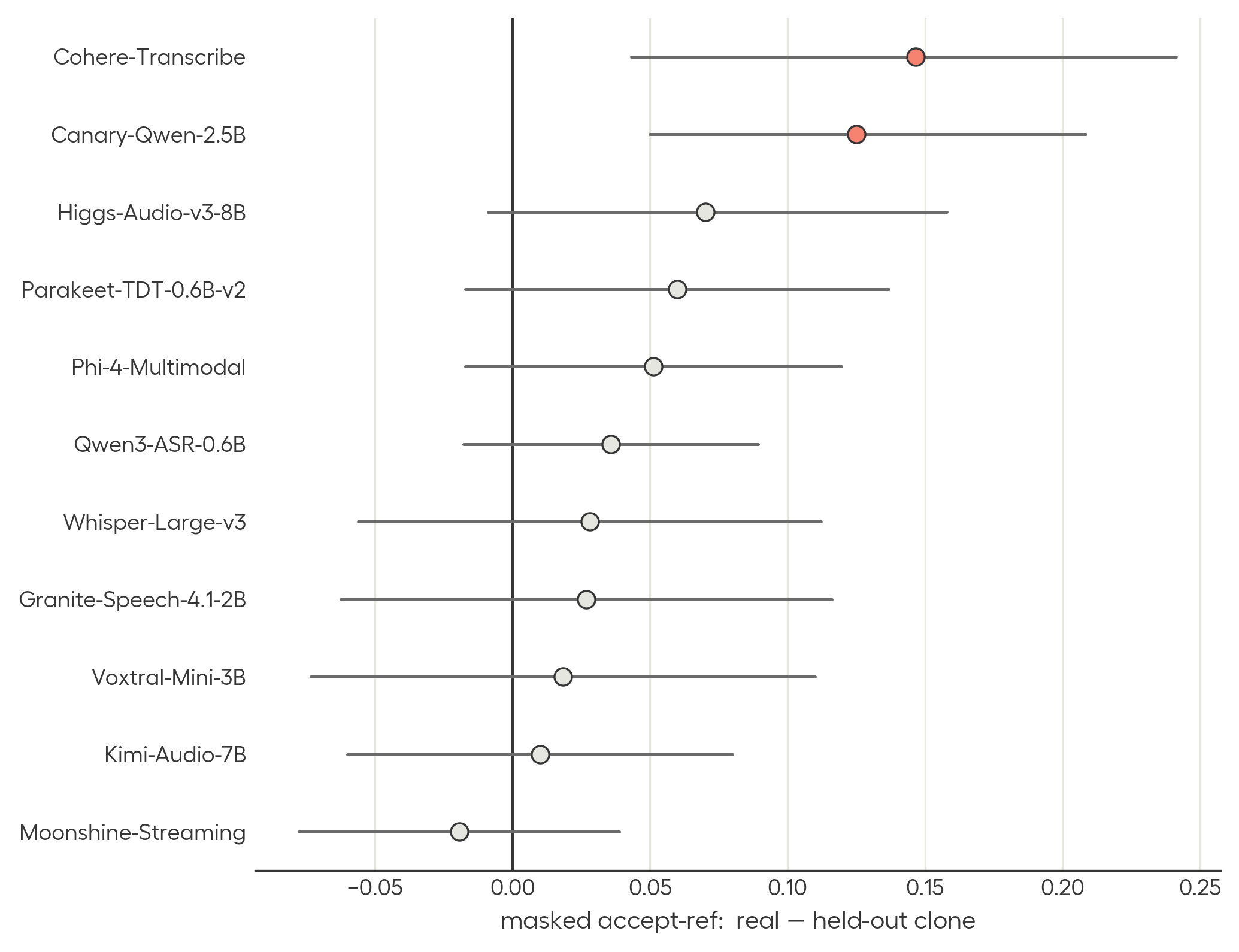}
\caption{Masked \textsc{accept-ref}, real VoxPopuli minus a register-matched held-out clone (fresh
post-cutoff parliament), same transcript, paired $95\%$ CI. Positive means the model reads a silenced
number from the benchmark voice but not a held-out voice; significant for Cohere and Canary (red).}
\label{fig:masking-voice}
\end{figure}

\begin{table}[tbp]\centering\small
\caption{Audio lift $\lambda(r)$ (Eq.~\ref{eq:lift}, nats/char) of the silenced number span by voice
condition ($115$ paired sentences passing the intelligibility gate in every clone condition). Diff columns: paired differences, bootstrap $95\%$ CIs;
\textbf{bold} marks CIs excluding zero. Whisper's lift \emph{rises} on clean TTS, so a drop in lift
is likely not a synthesis artifact; Qwen3's lift is negative in every condition, so its diffs do not indicate
recovery. Parakeet-TDT has no teacher-forced readout.}
\label{tab:nummask-lift}

\begin{tabular}{l cccc cc}
\toprule
& real & vox-clone & \textsc{ep-fresh} & generic & real$-$\textsc{ep-fresh} & real$-$generic \\
\midrule
Cohere-Transcribe     & $+1.52$ & $+1.26$ & $+0.92$ & $+0.54$ & $\mathbf{+0.60}$ {\scriptsize$[+0.30,+0.94]$} & $\mathbf{+0.98}$ {\scriptsize$[+0.66,+1.33]$} \\
Canary-Qwen-2.5B      & $+1.22$ & $+1.05$ & $+0.65$ & $+0.77$ & $\mathbf{+0.57}$ {\scriptsize$[+0.27,+0.90]$} & $\mathbf{+0.45}$ {\scriptsize$[+0.17,+0.71]$} \\
Granite-Speech-4.1-2B & $+0.12$ & $+0.15$ & $+0.16$ & $+0.01$ & $-0.04$ {\scriptsize$[-0.24,+0.17]$} & $+0.11$ {\scriptsize$[-0.11,+0.32]$} \\
Phi-4-Multimodal      & $+0.46$ & $+0.51$ & $+0.24$ & $+0.27$ & $\mathbf{+0.23}$ {\scriptsize$[+0.01,+0.50]$} & $\mathbf{+0.19}$ {\scriptsize$[+0.04,+0.35]$} \\
Higgs-Audio-v3-8B     & $+0.50$ & $+0.57$ & $+0.37$ & $+0.11$ & $+0.13$ {\scriptsize$[-0.00,+0.26]$} & $\mathbf{+0.39}$ {\scriptsize$[+0.23,+0.56]$} \\
\midrule
Whisper-Large-v3      & $+0.70$ & $+0.89$ & $+0.85$ & $+1.01$ & $-0.16$ {\scriptsize$[-0.37,+0.06]$} & $\mathbf{-0.31}$ {\scriptsize$[-0.47,-0.15]$} \\
Moonshine-Streaming   & $+0.31$ & $+0.37$ & $+0.25$ & $+0.15$ & $+0.06$ {\scriptsize$[-0.08,+0.21]$} & $+0.16$ {\scriptsize$[+0.00,+0.32]$} \\
Kimi-Audio-7B         & $+0.24$ & $+0.32$ & $+0.06$ & $+0.16$ & $+0.17$ {\scriptsize$[-0.00,+0.37]$} & $+0.08$ {\scriptsize$[-0.06,+0.22]$} \\
Qwen3-ASR-0.6B        & $-0.60$ & $-0.56$ & $-1.06$ & $-1.14$ & $\mathbf{+0.46}$ {\scriptsize$[+0.19,+0.75]$} & $\mathbf{+0.54}$ {\scriptsize$[+0.31,+0.76]$} \\
Voxtral-Mini-3B       & $-0.13$ & $+0.07$ & $-0.12$ & $-0.05$ & $-0.01$ {\scriptsize$[-0.17,+0.16]$} & $-0.08$ {\scriptsize$[-0.23,+0.07]$} \\
\bottomrule
\end{tabular}
\end{table}

\begin{table}[tbp]\centering\small
\caption{The opening-courtesy case study: rate at which the audible courtesy is present in the output. 
\textsc{truncated}: the audio is cut to the opener;
\textsc{attn-isolated} keeps the full-clip audio \emph{encoding} but restricts the decoder's attention
over it to the opener's frames. \textsc{translate}: the same audio decoded under an English$\to$Spanish translation
instruction. \textsc{full}: the entire clip.
-- marks models without
translation or attention-isolation capabilities.
}
\label{tab:trunc}

\begin{tabular}{l cccc}
\toprule
model & \textsc{truncated} & \textsc{attn-isolated} & \textsc{translate} & full \\
\midrule
Voxtral-Mini-3B       & 1.00 & 1.00 & 1.00 & 1.00 \\
Whisper-Large-v3      & 1.00 & 1.00 & 1.00 & 1.00 \\
Moonshine-Streaming   & 1.00 & 1.00 & --   & 1.00 \\
Qwen3-ASR-0.6B        & 0.94 & 0.95 & 1.00 & 1.00 \\
Kimi-Audio-7B         & 1.00 & 0.89 & --   & 0.89 \\
\midrule
Cohere-Transcribe     & 0.94 & 0.26 & -- & 0.00 \\
Granite-Speech-4.1-2B & 0.67 & 0.11 & 0.39 & 0.00 \\
Canary-Qwen-2.5B      & 0.83 & 0.05 & --   & 0.00 \\
Phi-4-Multimodal      & 0.83 & 0.89 & 0.61 & 0.00 \\
Higgs-Audio-v3-8B     & 0.83 & 0.42 & --   & 0.06 \\
Parakeet-TDT-0.6B-v2     & 1.00 & -- & --   & 0.00 \\
\bottomrule
\end{tabular}
\end{table}

\end{document}